# Independent Amplitude and Phase Modulation in Active Transmissive Metasurfaces via Complex Permittivity Tuning

Christopher M. Yi[1], Ruzan Sokhoyan[2], Juyoung Kim[2], Sungmin Hong[1], Jinwoo Song[1], Harry A. Atwater[2], and Min Seok Jang[1,2*]

[1]School of Electrical Engineering, Korea Advanced Institute of Science and Technology, Daejeon 34141, Republic of Korea

[2]Thomas J. Watson Laboratories of Applied Physics, California Institute of Technology, Pasadena, California 91125, USA.

*Corresponding author. Min Seok Jang: jang.minseok@kaist.ac.kr

## Abstract

Active metasurfaces are a promising platform for spatial light modulators (SLMs) in applications such as holography, beam shaping, and light detection and ranging (LiDAR), as they enable dynamic control over the amplitude and phase of light at subwavelength scales. Among these, transmissive metasurfaces offer a compact alternative to conventional SLMs, since they allow monolithic integration with chip-scale light sources. While independent amplitude and phase modulation have been previously demonstrated in active reflective metasurfaces, this important milestone has not been achieved in transmission due to the inherent challenges associated with the design of active transmissive metasurfaces. Here, we theoretically demonstrate an active transmissive metasurface that enables independent amplitude and phase modulation in the mid-infrared wavelength range. The proposed metasurface consists of high-Q cuboid germanium (Ge) Mie resonators integrated with graphene stripes. Utilizing the intraband and interband transitions of graphene, electrostatic gating enables modulation of the transmittance from near 0% to 5% at one spectral frequency and achieves 281° phase-only modulation at another frequency in the same device. To achieve both types of modulation at the same operating wavelength, we use the thermo-optic effect of Ge as a global tuning mechanism that spectrally aligns the resonant frequencies, enabling both transmittance modulation and phase-only modulation at different base temperatures. The proposed dual-mechanism architecture achieves transmittance modulation efficiency of ~100% and 282° phase-only modulation. Finally, we demonstrate the efficacy of phase-only modulation by individually addressing each unit cell to realize a beam-steering device with relative diffraction efficiencies over 90%. Our individually addressable active metasurface opens a route toward compact, dynamically reconfigurable metaphotonic devices.

## Introduction

Spatial light modulators (SLMs) enable dynamic wavefront control by modulating the amplitude or phase of light, underpinning applications such as holography, beam steering, light detection and ranging (LiDAR), and free-space optical communications.[1–4] However, conventional SLMs are limited by their bulky form factors, slow modulation speeds, and relatively large pixel sizes arising from liquid-crystal cross-talk and fabrication constraints, which restrict the achievable field of view and integration density. Active metasurfaces, planar arrays of subwavelength resonators integrated with tunable materials, have emerged as a transformative alternative that can address these limitations.[4–7] Utilizing metasurfaces operating in transmission could further reduce the SLM form factors since transmissive active metasurfaces can be monolithically integrated with on-chip light sources such as vertical-cavity surface-emitting lasers (VCSELs)[8,9] or photonic crystal surface-emitting lasers (PCSELs).[10,11]

Current strategies for implementing active metasurfaces to dynamically modulate amplitude and phase primarily leverage resonant phenomena to enhance light–matter interactions.[12–15] The nature of resonance, however, inherently couples the amplitude and phase responses, which complicates independent control of these two degrees of freedom.[12,16,17] While the coupling of amplitude and phase responses is inherent for both reflective and transmissive metasurfaces, there are additional challenges associated with transmissive configurations. In contrast to reflective metasurfaces, transmissive metasurfaces scatter light to both reflection and transmission ports, forming a two-port system. Prior research has shown that a lossless two-port single-resonance transmissive metasurface necessarily exhibits a transmission null near the resonance frequency.[17,18] This transmission null not only drives the transmittance to zero, but also introduces a $\pi$ phase jump at the null point, restricting the transmitted phase coverage to at most $\pi$. Additionally, when optical loss is introduced into a single-resonance two-port system, the transmission null becomes inaccessible, further limiting the achievable transmittance modulation efficiency and limiting phase modulation to $< \pi$.[17]

To circumvent the limitations of two-port single-resonance transmissive systems, previous works have introduced two-resonance architectures in which spectrally overlapping resonances satisfy the Kerker condition. This enables Huygens metasurfaces exhibiting unity transmittance and continuous $0-2\pi$ transmitted phase coverage in the spectral domain.[19–21] While various studies have demonstrated amplitude[22–24] and phase modulation[6,25–27] utilizing Huygens metasurfaces, achieving phase-only modulation with a single control parameter remains challenging due to the imbalanced response of the constituent resonances upon application of an external stimulus.[28] Introducing an additional control parameter can maintain the spectral overlap of the two resonances during phase modulation, but further complicates the device architecture even for achieving phase-only modulation alone.[26] A recent theoretical work has demonstrated that opening additional scattering ports to form a multi-port single-resonance system can overcome the physical limitation of transmissive two-port single-resonance systems, enabling nonzero transmittance values around the resonances while exhibiting a continuous 0–2$\pi$ transmitted phase coverage in the spectral domain.[17] Moreover, the mentioned work has demonstrated a lithium-niobate metasurface design that enables phase-only modulation of 300° through a single control parameter while maintaining an approximately constant transmittance.[17] However, the proposed metasurface design does not enable access to arbitrary phase–amplitude pairs upon application of an external stimulus.

Here, we extend the multi-port single-resonance framework to achieve independent amplitude and phase modulation of the transmitted light by using two independent external stimuli. First, we investigate an active metasurface design, consisting of high-Q germanium (Ge) Mie resonators integrated with graphene stripes. The key to decoupling amplitude and phase lies in the unique optical response of graphene, which supports both interband and intraband transitions. In the mid-infrared wavelength range, interband absorption dominates at low Fermi levels which governs the dissipative loss. As the Fermi level reaches half the photon energy ($\hbar\omega/2$), interband transitions are progressively suppressed by Pauli blocking, reducing the loss, while the intraband transitions become dominant and shift the resonance frequency. The operating wavelength of $\lambda$ = 6.5 μm (0.19 eV) is selected such that Pauli-blocking occurs near 0.1 eV, allowing a single electrostatic gating mechanism to exploit both transitions, thereby separately controlling the amplitude and phase. Leveraging each transition, the device achieves phase-only modulation in transmission, where the phase is dynamically tuned between 0° and 281° while transmittance stays almost constant. The proposed active metasurface also achieves a transmittance modulation efficiency approaching 100%, with the absolute transmittance varying from near 0% to ~5%.

To achieve both transmittance modulation efficiency of 100% and phase-only modulation at the same operating wavelength, we utilize the thermo-optic effect in Ge as a global tuning mechanism that spectrally aligns the resonance frequency, enabling both types of modulation at different base temperatures. The proposed dual-tuning architecture allows modulation of the transmittance from near 0% to 5%, and ~282° phase modulation with only a modest relative transmittance variation of 5%. Finally, by applying individually controlling gate voltages applied to each unit cell at the base temperature for phase-only modulation, the same device performs beam steering with relative diffraction efficiencies over 90% at first-order diffraction angles of 30°, 20°, and 15°. By independently controlling the temperature and applied bias, the proposed architecture enables independent amplitude and phase modulation at a single operating frequency, providing access to a broad range of complex transmission states.

## Results and Discussion

### Theoretical framework for independent amplitude and phase modulation

The capability for complex amplitude modulation is fundamentally governed by the constraints imposed on the complex transmission coefficient trajectory. Within the framework of temporal coupled-mode theory (TCMT), the complex transmission coefficient $t$ of an $n$-port system supporting a single resonance is expressed as[17,18,29]

$$t = t_{\mathrm{d}} + t_{\mathrm{r}} = t_{\mathrm{d}} + \frac{2\sqrt{\gamma_1\gamma_2}}{-i(\omega - \omega_{\mathrm{res}}) + \gamma}. \quad (1)$$

Equation (1) shows that the complex transmission coefficient t is determined by the interference of two pathways: a direct (non-resonant) pathway that bypasses the resonator, described by the direct transmission coefficient $t_{\mathrm{d}}$, and an indirect pathway mediated by the resonance, represented by $t_{\mathrm{r}}$. The interference between these direct and indirect pathways governs the amplitude and phase of the transmission coefficient. The resonance term is characterized by the resonance frequency $\omega_{\mathrm{res}}$, and the total decay rate of the resonant system $\gamma = \sum_{i=1}^{n}\gamma_i + \gamma_{\mathrm{d}}$, defined as the sum of all radiative decay rates $\gamma_i$ and the dissipative decay rate $\gamma_{\mathrm{d}}$. The subscripts 1 and 2 denote the input and output ports, respectively. Each port represents a radiative channel through which energy couples into or out of the resonance, and the coupling strength between the resonance and each port is characterized by the corresponding radiative decay rate. We adopt the convention of defining the port locations such that the radiative coupling coefficients are real-valued for convenience. As the resonant system exhibits a narrow spectral response, the direct transmission coefficient $t_{\mathrm{d}}$ can be assumed to remain constant within the spectral vicinity of the resonance, while the resonant term $t_{\mathrm{r}}$ traces a circular trajectory as a function of frequency with a radius of $R = \sqrt{\gamma_1\gamma_2}/\gamma$ in the complex transmission amplitude plane.

The physical configuration of the system imposes strict constraints on $t_{\mathrm{d}}$, which governs the position of this circular trajectory in the complex plane. In a lossless two-port single-resonance system, $t_{\mathrm{d}}$ is constrained by time-reversal symmetry, reciprocity, and energy conservation, forcing the complex transmission amplitude circle to pass through the origin, which necessarily exhibits a transmission null. In practical implementations, the presence of even minimal intrinsic loss increases the dissipative decay rate, reducing the radius of the resonant circle, precluding the trajectory from encircling or passing through the origin of the complex plane. This limitation typically results in a finite transmission minimum and restricts phase coverage to less than π. In contrast, the introduction of additional ports facilitates energy redistribution, thereby suppressing the direct transmission coefficient $t_d$ and relaxing the inherent physical constraints. This increased degree of freedom allows the resonant circle to be translated across the complex plane to either encircle or intersect the origin, even in the presence of loss, providing the basis for independent phase and amplitude modulation, respectively.

Active tuning of the metasurface modifies the complex permittivity of the active material, perturbing the resonant system and thereby primarily modifying the resonance frequency $\omega_{res}$ and the dissipative decay rate $\gamma_d$, within the TCMT framework. To illustrate how the resonance frequency and dissipative decay rate enables independent amplitude and phase modulation, we consider a four-port single-resonance system whose complex transmission amplitude circle in the frequency domain is centered at the origin of the complex plane. Under this condition, the transmission amplitude remains constant with 2π phase coverage in the frequency domain, establishing the basis for phase-only modulation. At the operating frequency $\omega_0$, which coincides or is in close vicinity of the resonance frequency $\omega_{res}$, tuning the resonance frequency $\omega_{res}$ moves the transmission coefficient along the circumference of the circle (thick solid curve in Figure 1a, progressing from purple to red as the tuning of $|\omega_{res} - \omega_0|$ increases), yielding near-2π phase modulation with negligible amplitude variations (Figure 1b, left panels). This ideal case assumes zero dissipative loss ($\gamma_d = 0$). When a finite $\gamma_d$ is introduced, as depicted by the thin solid curves in Figure 1a, the complex transmission amplitude circle shrinks and its center shifting away from the origin, as $t_d$ remains fixed. As the dissipative decay rate increases, the transmission dip deepens and amplitude variation increases, limiting phase-only modulation; however, a finite dissipative decay rate does not preclude phase-only modulation. Our full-wave simulations discussed below show that by optimizing the material and geometric parameters, we can modify $t_d$ to re-center the reduced circle at the origin, thereby restoring phase-only modulation even in a lossy system, albeit with diminished overall transmittance.

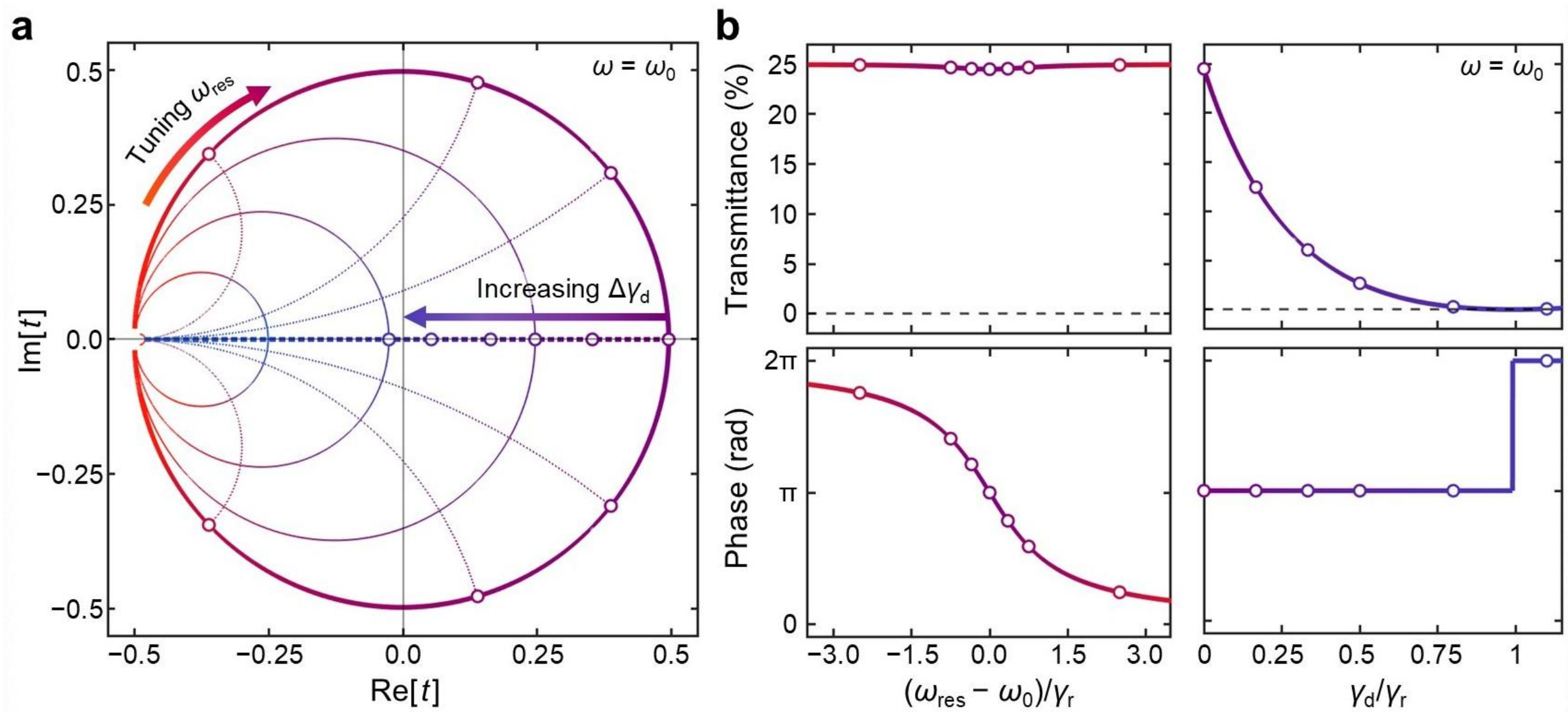

**Figure 1.** Theoretical framework for independent amplitude and phase modulation in a multi-port single-resonance system. (a) Complex transmission coefficient trajectories in the complex plane for a four-port single-resonance system. The thick solid curve (purple to red, increasing $|\omega_{res} - \omega_0|$) represents phase-only modulation in the ideal case ($\gamma_d = 0$), where the operating point traces the circumference of the transmission circle centered at the origin. Thin solid curves correspond to progressively increasing constant $\gamma_d$, showing reduced radius and transmittance. The thick dashed line (purple to blue, increasing $\gamma_d$) represents amplitude-only modulation in the ideal case ($\omega_{res} = \omega_0$), where the operating point moves radially towards the origin. Thin dashed lines correspond to $\omega_{res} \neq \omega_0$, illustrating deviation from the purely radial trajectory. (b) Transmittance (top) and phase (bottom) as a function of $(\omega_{res} - \omega_0)/\gamma_r$ for phase-only modulation (left) and $\gamma_d/\gamma_r$ for amplitude-only modulation (right). System parameters: $t_d = 0.5$, $\omega_0 = 1$, $\gamma_1 = \gamma_2 = \omega_0/1{,}000$, $\gamma_3 = \gamma_4 = \omega_0/200{,}000$.

While amplitude-only modulation does not strictly require the complex transmission amplitude circle to be centered at the origin, it necessitates that the circle encompasses the origin with its center located on the real axis, and is accessed by introducing a dissipative decay rate to the system. For simplicity, we illustrate amplitude-only modulation from the centered configuration discussed above. When the resonance frequency coincides with the operating frequency ($\omega_{res} = \omega_0$), increasing $\gamma_d$ contracts the radius of the complex transmission amplitude circle while the operating point remains on the real axis, driving it along a radial trajectory directly towards the origin (thick dashed line in Figure 1a, progressing from purple to blue as $\gamma_d$ increases). Along this trajectory, the phase remains constant while only the magnitude of the complex transmission coefficient $|t|$ varies, realizing amplitude-only modulation. Conventionally, the modulation efficiency is quantified in terms of the transmitted intensity, and we accordingly define it as $\eta_T = (T_{\max} - T_{\min})/T_{\max}$, where $T = |t|^2$ is the transmittance. As the radial contraction drives the trajectory through the origin, $T_{\min}$ reaches zero, yielding a transmittance modulation efficiency of 100% (Figure 1b, right panels). Notably, upon crossing the origin, a π phase jump naturally occurs, confirming that the trajectory precisely intersects the origin. When the operating frequency does not coincide with the resonance frequency ($\omega_0 \neq \omega_{\text{res}}$), the trajectory no longer follows a purely radial path (thin dashed lines in Figure 1a). Under this condition, the trajectory does not pass through the origin, precluding transmittance modulation efficiency of 100%, and simultaneously introduces an undesired phase variation. While modulation efficiency of 100% can be recovered by adjusting $t_{\text{d}}$ to re-position the transmission circle such that the trajectory intersects the origin, the accompanying phase variation persists, fundamentally compromising amplitude-only modulation. Furthermore, to achieve amplitude-only modulation with a transmittance modulation efficiency of 100% and a high maximum transmittance, operation in a regime where the dissipative decay rate is small relative to the radiative decay rates is advantageous. This advantage arises from the nonlinear dependence of transmittance on the dissipative decay rate. Because $\gamma_{\text{d}}$ appears in the denominator of the transmission coefficient and transmittance scales as $|t|^2$, the transmittance exhibits a nonlinear decay with diminishing sensitivity to $\gamma_{\text{d}}$ and exhibits a pronounced flattening near the origin of the complex plane (Figure 1b). Consequently, when the dissipative loss is small, even a small variation in $\gamma_{\text{d}}$ produces a substantial change in transmittance, yielding a large modulation efficiency. In contrast, when $\gamma_{\text{d}}$ is comparable to or exceeds the radiative decay rates, the trajectory lies within the flattened region of the transmittance curve, where a much larger change in $\gamma_{\text{d}}$ is required to achieve a comparable change in transmittance, making this regime less efficient for high-performance modulation.

These results underscore that for both phase-only and amplitude-only modulation regimes, employing a high-quality-factor resonance with minimal dissipative loss is preferable, as it provides a narrow linewidth that enables substantial phase modulation while simultaneously enhancing the sensitivity to loss variations for amplitude modulation, within a limited tuning range. In addition to this high-Q requirement, both regimes require the complex transmission amplitude circle to encircle the origin. Under this condition, ideal phase-only modulation necessitates that $\gamma_{\text{d}}$ remains constant while $\omega_{\text{res}}$ is varied; conversely, ideal amplitude-only modulation requires $\omega_{\text{res}}$ to be held fixed while $\gamma_{\text{d}}$ is modulated. In practice, if the two parameters are coupled, that is, if a single tuning mechanism simultaneously perturbs both $\omega_{\text{res}}$ and $\gamma_{\text{d}}$, the system traces a trajectory that is neither purely circumferential nor purely radial, significantly complicating the realization of independent modulation.

Beyond these amplitude-only and phase-only modulation regimes, if complete complex amplitude modulation can be achieved, any desired combination of amplitude and phase becomes independently addressable. This requires the transmission coefficient to access a two-dimensional region of the complex plane that encompasses the origin. The trajectory must be able to encircle the origin for full 2π phase modulation while also intersecting the origin for amplitude-only modulation with transmittance modulation efficiency of 100%. However, a single control parameter traces only a one-dimensional trajectory, a curve, in the complex plane, which inherently cannot cover a two-dimensional area. Therefore, two control parameters are essential to span the required region. By independently tuning $\omega_{\mathrm{res}}$ to sweep the phase along the circumference and $\gamma_{\mathrm{d}}$ to modulate the radius, the full complex plane region surrounding the origin can be covered, thereby enabling fully independent amplitude and phase modulation within a single device.

### A proof-of-concept metasurface design

Despite this theoretical potential, realizing independent phase and amplitude modulation remains a non-trivial task, as integrating both functionalities into a single platform presents a formidable challenge owing to their mutually exclusive operating conditions. To reconcile these disparate requirements, we demonstrate a metasurface that achieves both modulation regimes by utilizing graphene gating as a versatile tuning mechanism that can access both $\omega_{res}$ tuning and $\gamma_d$ tuning within distinct Fermi level regimes. With graphene alone, however, the inevitable resonance frequency shift precludes achieving both modulation types at the same operating frequency. To overcome this constraint, we further introduce the thermo-optic tuning of germanium (Ge) as an additional tuning mechanism, demonstrating transmittance modulation efficiency of 100% and near-2π phase-only modulation at a fixed frequency.

The proposed active transmissive metasurface is designed to facilitate dynamic complex amplitude modulation through the integration of graphene and Ge as tunable constituent materials. As illustrated in the schematics (Figure 2a, b), the metasurface comprises a periodic array of cuboid Ge resonators interconnected by thin Ge bars, a graphene–$HfO_2$–graphene capacitor structure, and a silica ($SiO_2$) substrate. Each constituent layer is engineered to fulfill a specific functional role in enabling complex amplitude modulation. Our architectural framework is based on a multi-port single-resonance system, where additional diffraction channels are intentionally introduced on the reflection side to maintain a multi-port system with a single-port configuration in transmission. To open these diffraction ports while preserving a single-port in transmission, the periods in the *x* and *y* directions are chosen to satisfy the condition $\lambda_0/n_{sub} < P_x = P_y < \lambda_0$. The $SiO_2$ substrate, having a higher refractive index than air, facilitates the opening of diffraction channels at the reflection side, enabling the complex transmission amplitude circle to encircle the origin and thereby providing 2π phase coverage in the spectral domain.

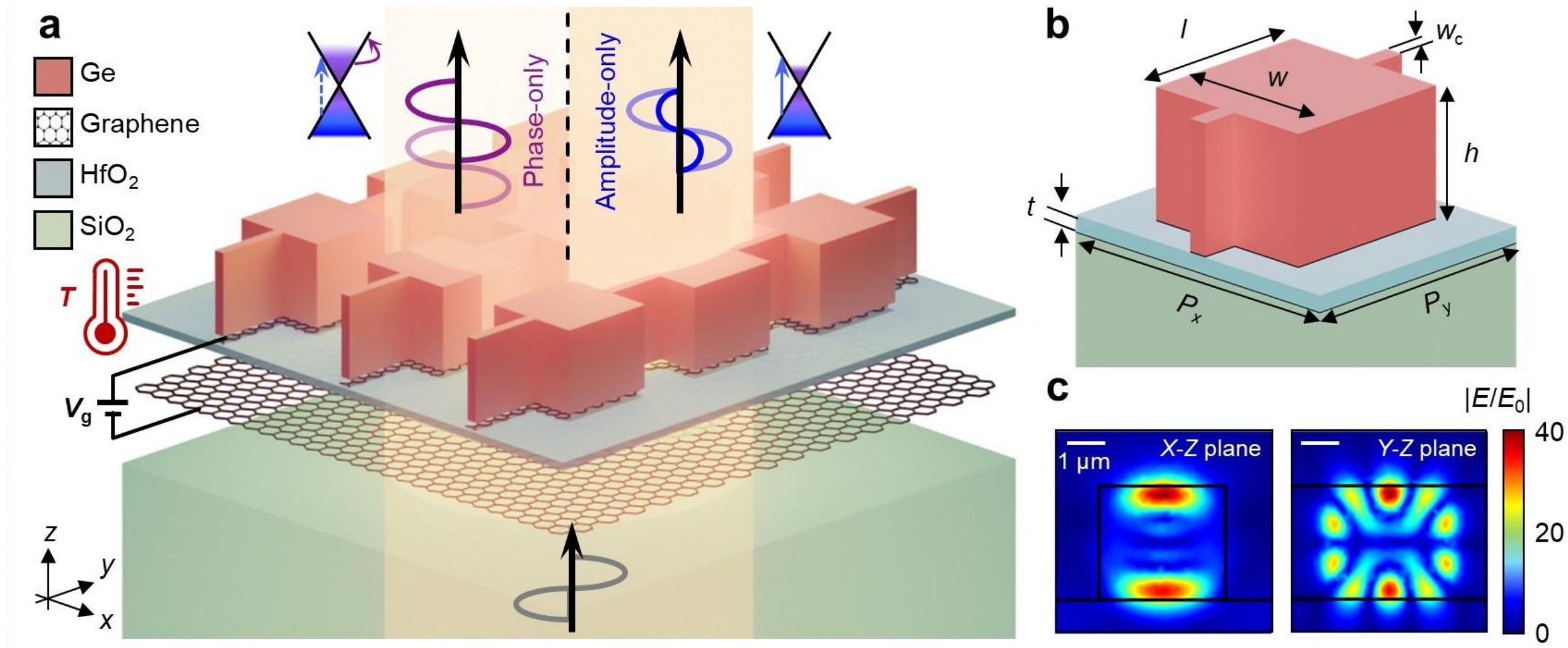

**Figure 2.** Design of the proposed active transmissive metasurface. (a) Schematic of the transmissive active metasurface integrated with graphene and germanium (Ge) for dynamic complex amplitude modulation. (b) Unit-cell geometry showing cuboid Ge resonators, interconnecting Ge bars, and graphene-$HfO_2$-graphene capacitors on $SiO_2$ substrate. $P_x$ and $P_y$ denote the periodicity in the $x$ and $y$ direction. $w$, $l$, $h$, $w_c$, $t$ represents the width, length, height of the Ge pillar, the width of the connecting bar, and the thickness of the $HfO_2$ layer, respectively. (c) Normalized cross-sectional electric field distributions in $X$-$Z$ ($y = 0$) and $Y$-$Z$ ($x = 0$) planes obtained from single unit cell simulation, showing dotriacontapole (32-pole) Mie resonance supported by the cuboid Ge resonator.

To maximize the modulation dynamic range, we employ cuboid Ge resonators that support a high-$Q$ dotriacontapole (32-pole) Mie resonance (Figure 2c).[30] Active modulation is achieved by integrating graphene, an exceptional mid-infrared material whose unique linear electronic dispersion enables substantial conductivity modulation via electrostatic gating. While many active metasurfaces rely on patterned graphene to excite plasmons for enhanced light–matter interaction, the highly confined electric fields characteristic of the dotriacontapole resonance enables strong modulation without relying on graphene plasmons. The graphene layer is placed beneath the Ge pillar and conforms to the Ge geometry due to the etching process. The cuboid Ge resonators are interconnected along the $y$-direction by thin Ge bars, under which the graphene layer extends continuously to maintain electrical continuity. This configuration enables multi-gate biasing along the $x$-axis for one-dimensional beam steering without requiring individual pixel addressing, which naturally isolates neighboring unit cells along the $x$-direction. Notably, the interconnecting bars are sufficiently thin relative to the cuboid Ge pillars so as to preserve the resonant field profile.

To gate the graphene, a thin $HfO_2$ layer is employed as a high-$k$ dielectric capacitor within a graphene-$HfO_2$-graphene capacitor structure. This architecture enables Fermi-level tuning up to 1.0 eV within a feasible voltage range without dielectric breakdown (Figure S1).[31,32] Assuming the graphene is initially intrinsic, grounding the bottom layer while applying a voltage to the top layer induces symmetric *n*-type and *p*-type doping, yielding identical absolute Fermi levels ($|E_F|$) in both layers. This configuration doubles the interaction with the resonant system compared with a single graphene layer configuration, resulting in a larger resonance shift (Figure S2a). Electrostatic simulations confirm that the vertical electric field distribution conforms to a parallel-plate capacitor model. In regions where the top graphene is patterned, the field enables Fermi-level tuning, whereas regions where the top graphene has been removed remain intrinsic ($E_F$ = CNP) (Figure S2b).

In contrast to the pixel-level electrical addressing of graphene, thermo-optic tuning is utilized as a global tuning mechanism that can be implemented by uniformly heating the entire metasurface using an external heater. As the electromagnetic field is predominantly confined within the Ge resonators, the thermo-optic response of the system is dominated by Ge, with graphene also contributing through its temperature-dependent conductivity.

**Independent amplitude and phase modulation via a single tuning mechanism at distinct wavelengths**

While $\omega_{res}$ tuning and $\gamma_d$ tuning impose distinct physical requirements on the active material, graphene's electronic transitions enable a single tuning mechanism to access transmittance modulation efficiency of ~100% and phase-only modulation within distinct Fermi level regimes. This is because graphene's optical response is governed by the interplay between interband and intraband transitions, which exert contrasting influences on the real and imaginary parts of its conductivity. The crossover between these processes is dictated by the Pauli blocking threshold ($E_F = \hbar\omega/2$), effectively partitioning graphene's modulation capabilities into two distinct regimes. To fully exploit both mechanisms, we operate in the mid-infrared spectral range at a wavelength of 6.5 μm (0.191 eV). In the mid-infrared wavelength range, the Pauli blocking starts at $E_F$ = 0.095 eV and interband transitions are progressively suppressed above this threshold, becoming almost completely suppressed around 0.2 eV.

Previous graphene-based active metasurfaces have primarily targeted either amplitude or phase modulation, without independent control of the two.[12,15,33,34] Likewise, active metasurfaces based on other tuning mechanisms, such as liquid crystals,[6,28] and phase-change materials,[22,25] and electro-optic materials,[17,35,36] have modulated amplitude and phase in a coupled manner, where tuning one quantity inevitably alters the other. In contrast, two key features of our design enable access to both amplitude-only and phase-only modulation in a single device. First, the multi-port single-resonance system supported by the Ge resonator not only provides the flexibility to position the complex transmission amplitude circle at the origin, but also sustains a high-Q resonance that achieves comparable radiative and dissipative decay rates, amplifying the device sensitivity to graphene's modulation. Second, the use of $HfO_2$ as a high-*k* dielectric enables Fermi-level tuning up to 1.0 eV and operating in the mid-infrared wavelength range places the Pauli blocking threshold at a sufficiently low Fermi level, ensuring clearly separated interband and intraband regimes with sufficient tuning range for independent $\omega_{res}$ and $\gamma_d$ control.

The physical connection between graphene's conductivity and its role in the resonant system can be understood through first-order perturbation theory. Active modulation modifies the conductivity of the graphene layer, perturbing the resonant system and shifting the complex eigenfrequency of the resonant mode as[37]

$$\Delta\widetilde{\omega} = -\frac{\omega_{res}}{2}\frac{\int dv \Delta\epsilon(\boldsymbol{r})|\boldsymbol{E}(\boldsymbol{r})|^2}{\int dv \epsilon(\boldsymbol{r})|\boldsymbol{E}(\boldsymbol{r})|^2} \quad (2)$$

where $\widetilde{\omega} = \omega_{\mathrm{res}} - i\gamma$ is the complex eigenfrequency of the mode, and $\mathbf{E}(\mathbf{r})$ is its electric field. $\varepsilon(\mathbf{r})$ and $\Delta\varepsilon(\mathbf{r})$ are the permittivity distribution and its change induced by the external stimulus, respectively. The integrals are taken over the unit cell of the metasurface. The graphene permittivity is expressed as $\varepsilon = \varepsilon_0 + i\sigma/(\omega t_g)$, where $\sigma$ is the surface conductivity, $\omega$ is the angular frequency, and $t_g = 0.34$ nm is the effective thickness of the graphene layer. From this relation, the imaginary part of the conductivity Im($\sigma$) modifies the real part of the permittivity, shifting the resonance frequency $\omega_{\mathrm{res}}$, whereas the real part Re($\sigma$) contributes to the imaginary part of the permittivity, changing the dissipative decay rate $\gamma_d$, which reflects the material loss. To quantitatively evaluate these relationships, the parameters ($\gamma_d$ and $\omega_{\mathrm{res}}$) were extracted using TCMT and Fano fitting, confirming that Re($\sigma$) directly governs $\gamma_d$, while Im($\sigma$) governs $\omega_{\mathrm{res}}$ (see Supporting Information Section S2). In the low-Fermi-level regime ($E_F < 0.1$ eV), interband transitions dominate, producing a substantial variation in Re($\sigma$) while Im($\sigma$) remains nearly constant (Figure 3a, b). This selectively modulates $\gamma_d$ while keeping $\omega_{\mathrm{res}}$ stable, providing the mechanism for amplitude-only modulation. In the high-Fermi-level regime ($E_F > 0.2$ eV), interband transitions are suppressed by Pauli blocking as intraband transitions become dominant, Im($\sigma$) increases monotonically, while Re($\sigma$) remains relatively stable (Figure 3a, b). This selectively modulates $\omega_{\mathrm{res}}$ while keeping $\gamma_d$ stable, providing the mechanism for phase-only modulation.

Leveraging the substantial Re($\sigma$) variation in the low-Fermi-level regime, the complex transmission amplitude circle can be driven to pass directly through the origin of the complex plane (Figure 3c). As $E_F$ approaches the charge neutrality point (CNP), the increased $\gamma_d$ reaches a magnitude comparable to the radiative decay rates, triggering a critical transition. At an operating wavelength of $\lambda = 6.509$ μm (Figure 3d), the system satisfies the condition for a transmission null, enabling transmittance modulation efficiency of 97% ($T_{\mathrm{min}} = 0.16\%$, $T_{\mathrm{max}} = 5.6\%$) (Figure 3e). For $E_F < 0.1$ eV, Im($\sigma$) remains nearly constant, and the corresponding transmission coefficient trajectory follows a radial path characteristic of ideal amplitude-only modulation (Figure 3c, dashed line, Figure S9a). However, operating strictly within this limited tuning range restricts the maximum achievable transmittance, as the small variation in Re($\sigma$) produces only a modest change near the origin.

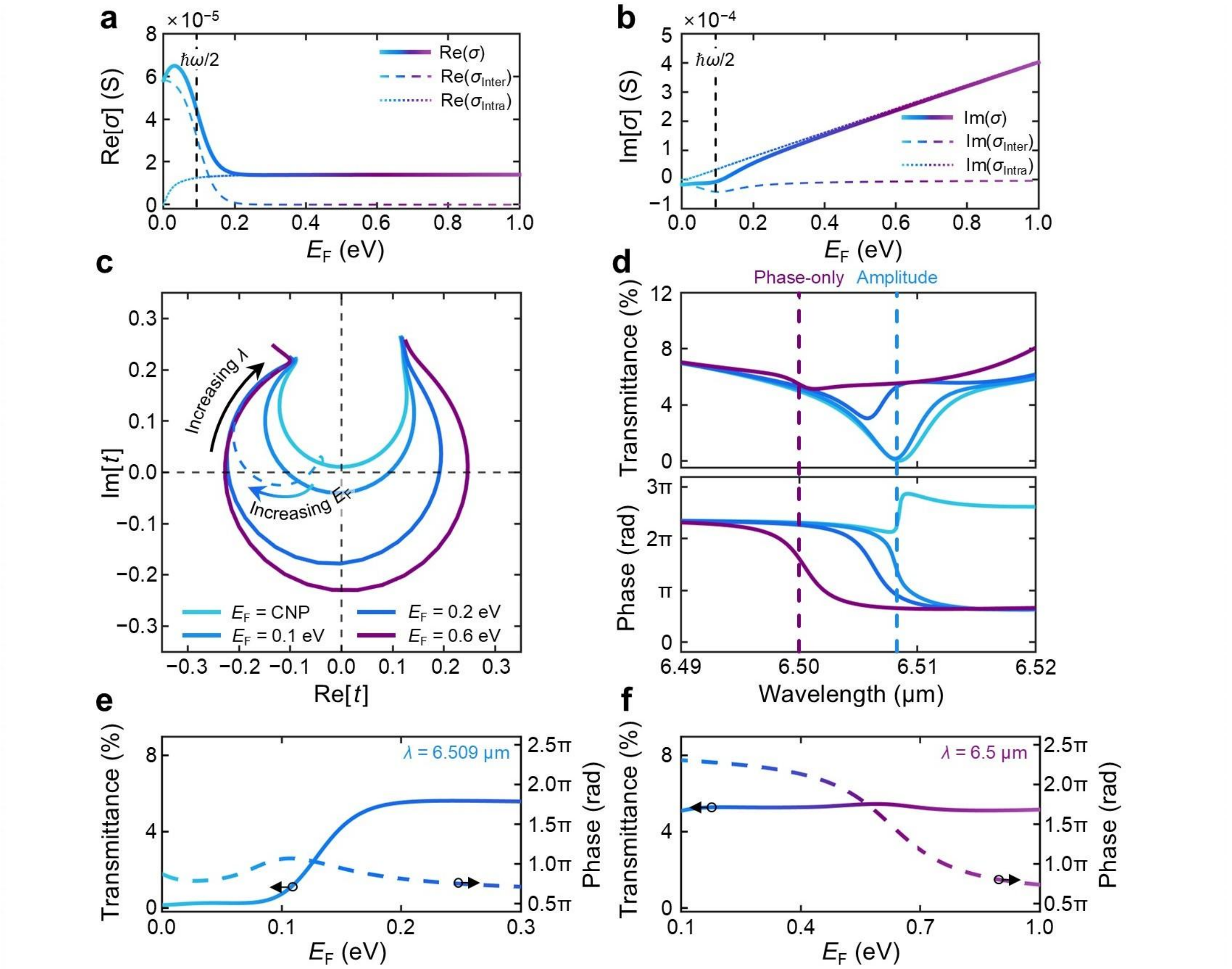

**Figure 3.** Graphene gating as a tuning mechanism for independent amplitude and phase modulation. (a, b) Real (a) and imaginary (b) parts of graphene's complex optical conductivity as a function of Fermi level ($E_F$) at $\lambda = 6.5$ μm. Solid lines represent the total conductivity, while dashed and dotted lines correspond to the interband and intraband contributions, respectively. The vertical black dashed line indicates the Pauli blocking threshold. ($E_F = \hbar\omega/2$) (c) Complex transmission coefficient trajectories in the frequency domain at different Fermi levels, showing the critical transition through the origin at low $E_F$ (amplitude modulation) and the relatively stable trajectory at high $E_F$ (phase-only modulation). Dashed lines indicate the trajectories at $\lambda = 6.509$ μm as $E_F$ is varied. (d) Spectral transmittance (top) and phase (bottom) for the corresponding trajectories showing amplitude modulation at $\lambda = 6.509$ μm (blue dotted line) and phase-only modulation at $\lambda = 6.500$ μm (purple dotted line). (e) Transmittance as a function of Fermi level at $\lambda = 6.509$ μm, achieving a modulation depth of nearly 100%. (f) Transmittance and phase as a function of Fermi level at $\lambda = 6.500$ μm, achieving 281° phase modulation while maintaining a stable transmittance of 5% with transmittance variation below 7%. The operating temperature is $T$ = 293 K. The geometric parameters are $P_x$ = 6.434 μm, $P_y$ = 5.845 μm, $w$ = 3.761 μm, $l$ = 3.754 μm, $h$ = 3.379 μm, $w_c$ = 0.481 μm, $t$ = 0.087 μm.

To enhance the maximum transmittance, the Fermi level is increased toward 0.2 eV to decrease the dissipative decay rate and enlarge the complex transmission amplitude circle. In this extended range (0.1 eV < $E_F$ < 0.2 eV), however, the subsequent variation in Im($\sigma$) induces a shift in $\omega_{res}$, bending the transmission coefficient into a curved path. Consequently, the substantial increase in transmittance in this regime is not solely attributed to the reduction in $\gamma_d$ but is also driven by this resonance shift. As $\omega_{res}$ moves away from the operating frequency $\omega_0$, the resonant contribution diminishes and the transmission coefficient rapidly converges toward $t_d$, sharply increasing the transmittance. As the maximum transmittance is increased for amplitude modulation, this introduces an undesired phase variation accompanying amplitude modulation, limiting amplitude-only modulation.

In the high-Fermi-level regime, the substantial Im($\sigma$) variation enables $\omega_{res}$ tuning for phase-only modulation. However, active graphene tuning can simultaneously alter the radiative decay rate, causing the complex transmission amplitude circle to change its size and complicating the achievement of ideal phase-only modulation. Furthermore, the high-Q resonance is exceptionally sensitive to geometric dimensions, adding another layer of complexity to the forward design. While gradient-based inverse design has been employed for free-form topology optimization of nanophotonic structures,[38,39] the advantages of gradient-descent optimization can be extended to fixed-form parametric structures in high-Q resonant systems. Accordingly, we employ gradient-descent optimization to co-optimize the geometric dimensions and the operating frequency to maximize the phase-only modulation performance. The figure of merit (FoM) for phase-only modulation is defined to simultaneously maximize the phase coverage and transmittance while minimizing the amplitude variation along the complex transmission trajectory (see Supporting Information Section S4). While the geometric parameters are specifically optimized for phase-only modulation in the high-Fermi-level regime, the high-Q resonance system naturally accommodates amplitude modulation in the low-Fermi-level regime ($E_F$ < 0.2 eV) as the radiative decay rate has a magnitude comparable to the graphene-induced dissipative loss. Through this optimization, we identify a robust operating point at $\lambda$ = 6.5 µm, achieving a phase modulation of 281° while maintaining a stable transmittance of ~5% with transmittance variation below 7% (Figure 3f, Figure S9b). The system sustains high-Q resonances (1,000 < $Q$ < 1,800) even in the presence of graphene-induced loss, providing the narrow linewidth necessary to achieve near-2π phase coverage within graphene's limited tuning range.

These results demonstrate that both amplitude-only and phase-only modulation can be effectively accessed by utilizing graphene as a single tuning mechanism, albeit at distinct spectral frequencies due to the resonance frequency shift.

**Dual tuning mechanism for independent amplitude and phase modulation at a fixed wavelength**

As demonstrated in the previous section, graphene was established as a viable tuning mechanism capable of independently accessing both amplitude and phase. However, the spectral shift associated with resonance tuning prevents these two modulations from being realized at a single operating frequency using only a single control parameter. To realize both modulations at a single operating wavelength, an additional control parameter is required to compensate for this spectral mismatch. Therefore, we utilize the thermo-optic effect of Ge as an additional tuning mechanism that exclusively controls $\omega_{res}$ without perturbing the decay rate. By spectrally aligning the resonance position via Ge thermal tuning, we allow local graphene gating to perform independent amplitude and phase modulation at a fixed wavelength. Furthermore, because the electromagnetic field is highly confined within the Ge resonators, Eq. (2) indicates that even a slight variation in the refractive index—corresponding to a temperature change of only 13 K—induces a significant resonance shift sufficient to compensate for the spectral mismatch caused by graphene tuning.

The metasurface was re-optimized using the same phase-only modulation figure of merit at $T = 310$ K, using the geometric parameters of the previously optimized metasurface as an initial point. In this configuration, temperature acts as a global tuning mechanism that switches amplitude and phase modulation. Specifically, the device operates in the amplitude modulation regime at low temperatures and transitions to the phase-only modulation regime at elevated temperatures. The optimized results are shown in Figure 4, where a phase modulation of 282° is achieved with the transmittance maintained near 5% (with a transmittance variation below 5%) along Path 1. By lowering the global temperature of the metasurface, the operating wavelength of amplitude modulation shifts to overlap with the phase-only modulation wavelength. This overlap allows the transmittance to approach a near-zero state ($T_{min} < 10^{-4}\%$, $T_{max} = 6.3\%$), effectively crossing the origin of the complex plane, showing a transmittance modulation efficiency of 100% (Path 2 in Figure 4). Although both modulations now occur at the same frequency, the underlying physical mechanism of graphene gating remains unchanged: amplitude modulation is dominant in the low Fermi level regime, while phase-only modulation governs the high Fermi level regime.

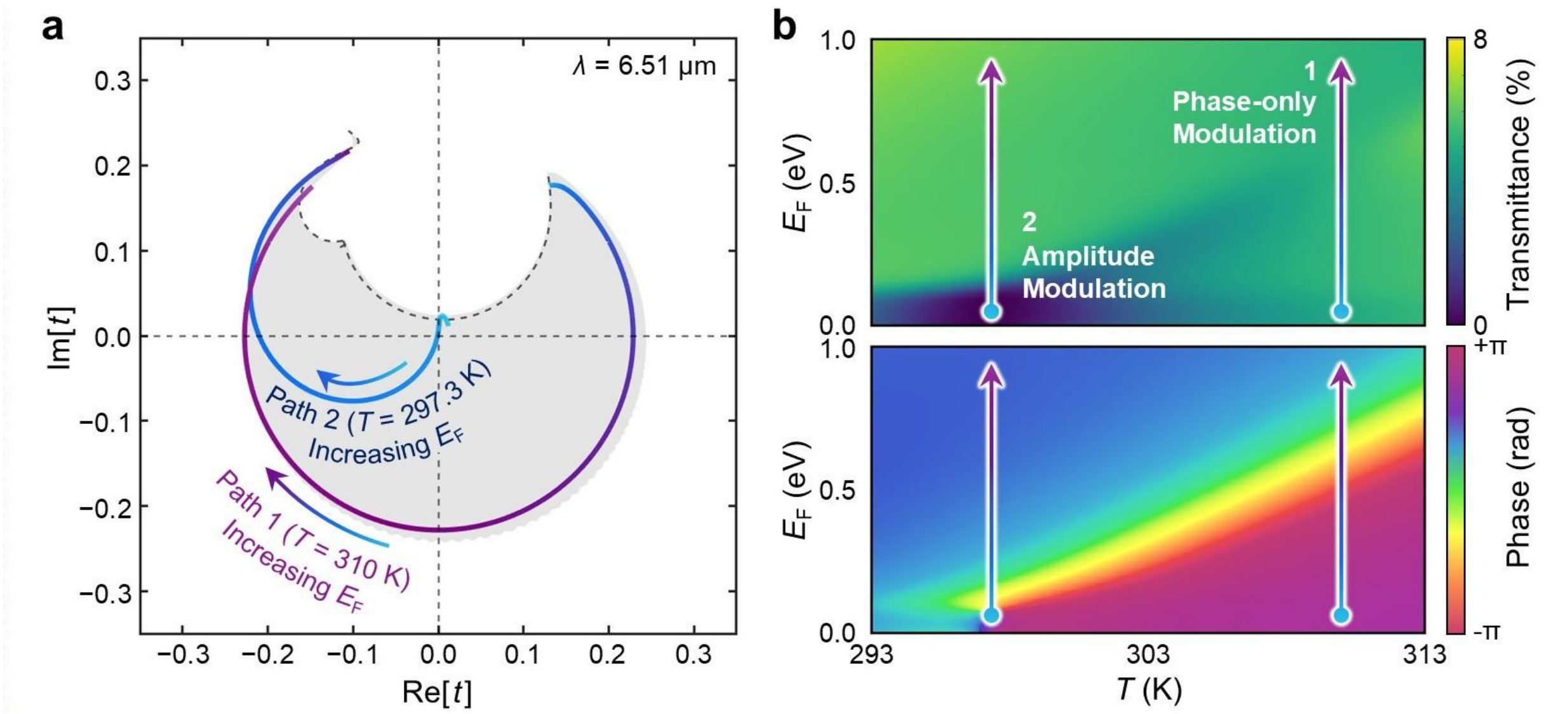

**Figure 4.** Independent amplitude and phase modulation at a fixed wavelength via dual tuning mechanisms. (a) Complex transmission coefficient trajectories in the complex plane at $\lambda = 6.51$ μm, showing amplitude and phase modulation paths driven by varying the Fermi level at fixed temperature. Path 1 ($T = 310$ K, varying $E_F$) demonstrates phase-only modulation with 282° phase modulation while maintaining a stable transmittance of 5% with transmittance variation below 5%. Path 2 ($T = 297.3$ K, varying $E_F$) achieves transmittance modulation efficiency of ~100%, driving the transmittance from 5% to near-zero. The shaded grey area indicates the accessible region in the complex plane spanned by both control parameters, confirming fully independent amplitude and phase modulation. (b) Transmittance (top) and phase (bottom) mapped across the $T$–$E_F$ parameter space at $\lambda = 6.51$ μm, with the two modulation paths overlaid. The geometric parameters are $P_x = 6.415$ μm, $P_y = 5.858$ μm, $w = 3.761$ μm, $l = 3.754$ μm, $h = 3.377$ μm, $w_c = 0.481$ μm, $t = 0.093$ μm.

Finally, while the thermal tuning mechanism in this work operates globally due to the high thermal cross-talk, several design approaches have shown localized thermal control of metasurfaces.[30,40] Realizing such control would require suppressing thermal cross-talk between neighboring elements, such as by increasing the gap or introducing pedestals, to sustain distinct local temperatures at a practical operating temperature. With local thermal control, thermo-optic tuning could serve as a spatially addressable phase-only modulation mechanism, as it tunes $\omega_{res}$ without perturbing the decay rates, while graphene gating serves as the amplitude modulation mechanism through $\gamma_d$ tuning. Since each mechanism independently governs $\omega_{res}$ and $\gamma_d$, controlling both parameters at each pixel would enable access to the entire complex transmission plane, providing an intuitive framework for spatially resolved complete complex amplitude modulation (see Supporting Information Section S5).

**Active beam steering via phase gradient**

The dual-tuning architecture enables both amplitude and phase modulation at a fixed operating frequency. While amplitude modulation can be directly used for transmission modulation without requiring spatial control, phase-only modulation enables wavefront shaping, which requires spatially varying control at the unit-cell level. Accordingly, to validate this capability, we design a one-dimensional (1D) beam-steering device.[6,41–44] In this architecture, independent temporal control of the phase response at the unit-cell level enables the synthesis of reconfigurable wavefronts. The operational performance and supercell configurations are illustrated in Figure 5. To implement dynamic beam steering, a continuous 2π phase distribution is discretized into $N$ distinct phase levels within a supercell, resulting in a uniform phase step of $\Delta\varphi = 2\pi/N$ between adjacent unit cells. The interconnecting Ge bars ensure electrical connectivity for the graphene layers along the $y$-direction, permitting temporal reconfiguration of the phase response for each unit row along the $x$-direction. By applying independent Fermi levels (ranging from 0.1 to 1.0 eV) to individual bars, a linear phase gradient is established to govern wavefront shaping in accordance with the generalized Snell's law. Modulation of the phase step $\Delta\varphi$ between adjacent unit cells dynamically adjusts the effective supercell period, defined as $\Lambda_{\text{supercell}} = N \times \Lambda$ (where $\Lambda$ is the unit-cell periodicity), thereby shifting the first-order diffraction angle θ according to the grating equation, which is physically consistent with the generalized Snell's law:

$$sin\theta = \frac{\Delta\phi}{2\pi}\frac{\lambda}{d} = \frac{\lambda}{\Lambda_{\text{supercell}}}. \quad (3)$$

However, several non-ideal factors, specifically phase discretization and near-field coupling between neighboring unit cells, can degrade the diffraction efficiency.[45] To mitigate these effects, the Fermi levels of each unit cell are optimized via a gradient-descent algorithm, using the geometric parameters obtained from the phase-only modulation optimization in Figure 4. The relative efficiency of the first-order diffraction serves as the figure of merit for this optimization.

Figure 5a–c presents the results for a supercell with $N = 2$. Due to structural symmetry, the $+1^{st}$ and $-1^{st}$ order diffraction efficiencies are identical, allowing the device to function as a beam splitter. While an initial phase difference of π yields a diffraction efficiency of only 60% at the target frequency (see Supporting Information Section S6), gradient-descent optimization identifies a refined phase difference of 0.65π, achieving a relative efficiency of 99.9% at a steering

angle of 30.5°, with the corresponding field profile presented in Figure 5c. In contrast, the optimized results for $N = 3$ (Figure 5d–f) and $N = 4$ (Figure 5g–i) converge towards the ideal phase gradients of 2π/3 and 2π/4, respectively, achieving high relative efficiencies of 96.9% at 19.8° ($N = 3$) and 93.1% at 14.7° ($N = 4$), with corresponding field profiles shown in Figure 5f, i. These results indicate that at lower quantization levels (e.g., $N = 2$), the large phase steps between adjacent cells exacerbate near-field coupling effects, necessitating significant deviations from the ideal linear gradient to compensate for coupling-induced phase errors.

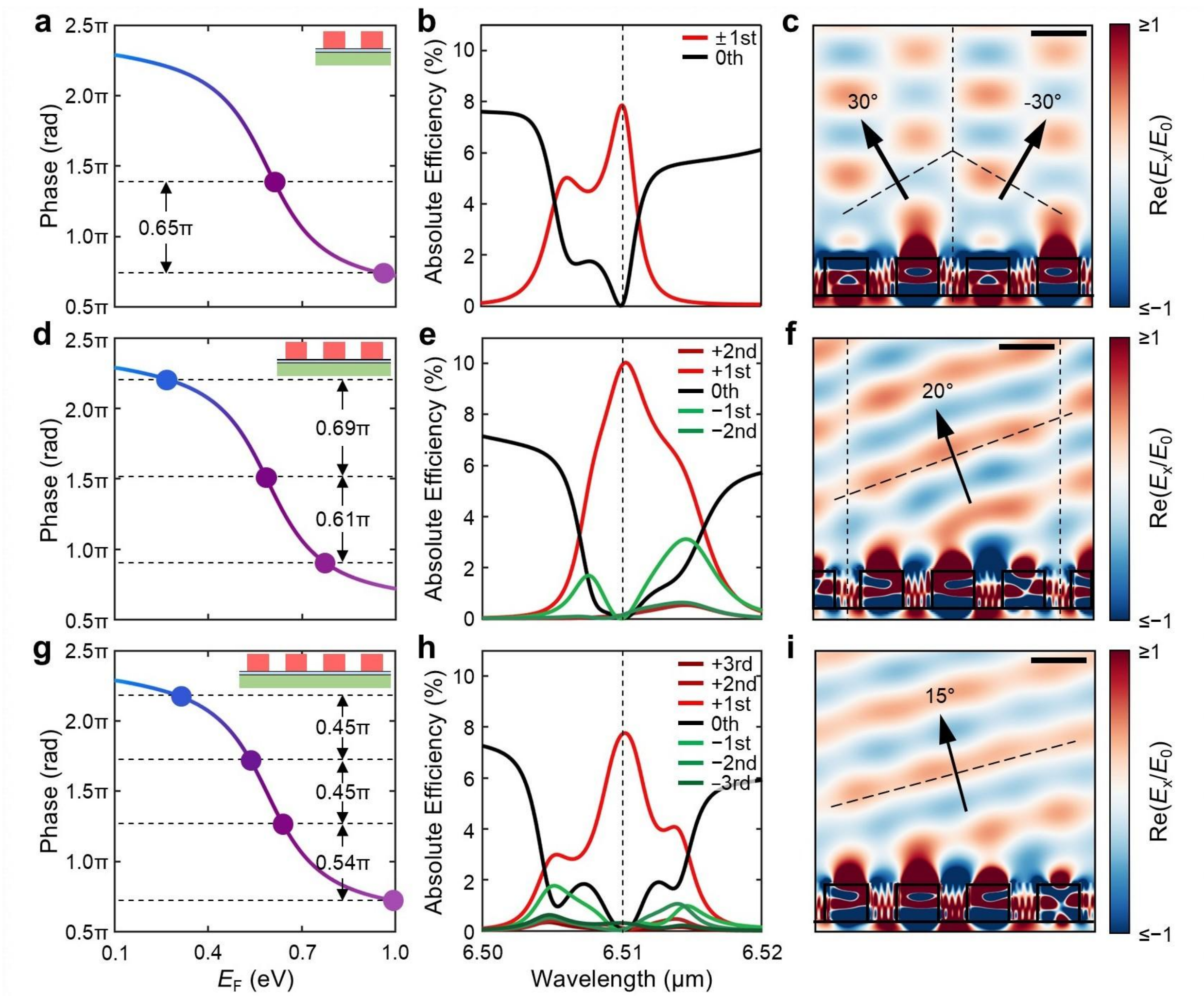

**Figure 5.** Active beam-steering metasurface via phase gradient (a, d, g) Phase modulation as a function of Fermi level at $\lambda = 6.51$ μm and $T = 310$ K. Optimized Fermi levels (dots) to maximize relative diffraction efficiency for supercells with (a) $N = 2$ (0.636 eV, 0.972 eV), (d) $N = 3$ (0.270 eV, 0.583 eV, 0.772 eV), and (g) $N = 4$ (0.318 eV, 0.536 eV, 0.639 eV, 0.991 eV). (b, e, h) Absolute efficiency of each diffraction order at the resonant wavelength when (b) two, (e) three, (h) four Fermi levels are applied. (c, f, i) Simulated electromagnetic field profile at the resonance frequency

$\lambda$ = 6.51 μm when (c) two, (f) three, (i) four Fermi levels are applied. All scale bars are 5 μm.

## Conclusions

In conclusion, we have demonstrated an active transmissive metasurface capable of independent amplitude and phase modulation through the integration of dual tuning mechanisms within a multi-port single-resonance framework. With graphene as a single tuning mechanism, we achieved transmittance modulation efficiency of ~100% and near-2π phase-only modulation within the same device at distinct spectral frequencies. To overcome the spectral limitation of graphene tuning alone, the thermo-optic effect of Ge was introduced as a secondary tuning mechanism, achieving transmittance modulation efficiency of ~100% and phase modulation of ~282° with transmittance variations suppressed below 5%. The efficacy of phase-only modulation is further validated through a beam-steering device with relative diffraction efficiencies exceeding 90%. While this work specifically utilizes graphene and Ge tuning, the underlying approach, in which each modulation degree of freedom is governed by a distinct physical process, is general and applicable to other material platforms. This architecture offers a foundational platform for next-generation active metaphotonics, including dynamic holography, advanced optical communications, and programmable wavefront engineering.

## Methods

*Electromagnetic simulation for structure optimization*

Full-wave electromagnetic simulations were performed using RETICOLO, a rigorous coupled-wave analysis (RCWA) solver.[46,47] In all optical simulations, a normally incident, linearly polarized plane wave illuminates the metasurface from within the substrate. The refractive index of Ge was obtained from Li[48] at $T = 293$ K, while the dispersion relations for $SiO_2$ and $HfO_2$ were obtained from Malitson[49] and Franta,[50] respectively. The surface conductivity of graphene was calculated using the random-phase approximation[51] with a carrier mobility of $\mu = 1{,}000$ $cm^2$ $V^{-1}$ $s^{-1}$. For temperature-dependent simulations, the thermo-optic coefficient of Ge ($dn$ / $dT = 4 \times 10^{-4}$ $K^{-1}$) was estimated from the refractive index values reported by Li[48] at 293 K and 350 K, assuming a linear temperature dependence over this range. The conductivity of graphene was also calculated as a function of temperature. The thermo-optic coefficients of $SiO_2$[52] and $HfO_2$[53] are approximately one order of magnitude smaller than that of Ge, and their refractive indices were therefore treated as temperature-independent in all simulations. The Fourier order required for RCWA to reproduce transmission coefficient spectra consistent with finite-element method (FEM) results was 10 for both the $x$- and $y$-directions. The optical responses of the optimized metasurfaces computed by RCWA were validated against a commercial FEM solver (COMSOL Multiphysics), showing excellent agreement. The transmission coefficient was extracted at a monitor placed at least $5\lambda$ from the Ge resonator to ensure far-field measurement. The phase was compensated by subtracting the propagation phase obtained from a reference simulation without the Ge resonator.

*Gradient-descent optimization*

Iterative gradient-descent optimization was driven by automatic differentiation (AD) using MEENT, a differentiable RCWA solver.[54] In both cases, the parameters were updated using the Adam optimizer.[55] For phase-only modulation, geometric parameters were optimized to maximize the figure of merit under both electro-optic (graphene Fermi level) and thermo-optic (Ge refractive index) tuning conditions. For the beam-steering device, the graphene Fermi levels of individual meta-atoms were optimized to maximize the first-order diffraction efficiency. In the supercell simulations, the Fourier orders along the $x$-direction were scaled proportionally to the supercell size. Specifically, the number of harmonics was doubled for $N = 2$, tripled for $N = 3$, and quadrupled for $N = 4$, to ensure numerical convergence commensurate with the enlarged periodicity.

**Author Contributions**

C.M.Y., R.S., and M.S.J conceived the research idea. C.M.Y. and S.H. conducted device optimization. C.M.Y. performed electromagnetic simulations. C.M.Y. and M.S.J. analyzed the data. J.S. contributed to the conceptual visualization and schematic illustration. M.S.J. supervised the project. C.M.Y. wrote the original draft, and R.S., J.K., H.A.A., and M.S.J. reviewed and edited it.

**Acknowledgment**

This research was supported by Culture, Sports and Tourism R&D Program through the Korea Creative Content Agency (KOCCA) grant funded by the Ministry of Culture, Sports and Tourism in 2024 (RS-2024-00332210), Basic Science Research Program through the National Research Foundation of Korea (NRF) funded by the Ministry of Science and Technology (RS-2022-NR070476, RS-2024-00416583, RS-2024-00414119).

**Conflict of Interest**

The authors declare that they have no known competing financial interests or personal relationships that could have appeared to influence the work reported in this paper.

# Supplementary Information for Independent Amplitude and Phase Modulation in Active Transmissive Metasurfaces via Complex Permittivity Tuning

Christopher M. Yi[1], Ruzan Sokhoyan[2], Juyoung Kim[2], Sungmin Hong[1], Jinwoo Song[1], Harry A. Atwater[2], and Min Seok Jang[1,2*]

[1]School of Electrical Engineering, Korea Advanced Institute of Science and Technology, Daejeon 34141, Republic of Korea

[2]Thomas J. Watson Laboratories of Applied Physics, California Institute of Technology, Pasadena, California 91125, USA.

*Corresponding author. Min Seok Jang: jang.minseok@kaist.ac.kr

**This PDF file includes:**

Supplementary Notes S1 to S9

Supplementary Figures S1 to S18

**S1. Graphene-$HfO_2$-graphene capacitor**

To electrostatically gate the graphene in the transmissive configuration, we employ a graphene–$HfO_2$–graphene capacitor structure, in which both graphene layers serve as electrodes and $HfO_2$ functions as the gate dielectric. To calculate the required voltage to reach the Fermi level of 1.0 eV, we use the parallel-plate capacitor model[1] with a $HfO_2$ dielectric constant of 22 and a thickness of 93 nm, resulting in a gate voltage of approximately 55 V (Figure S1).

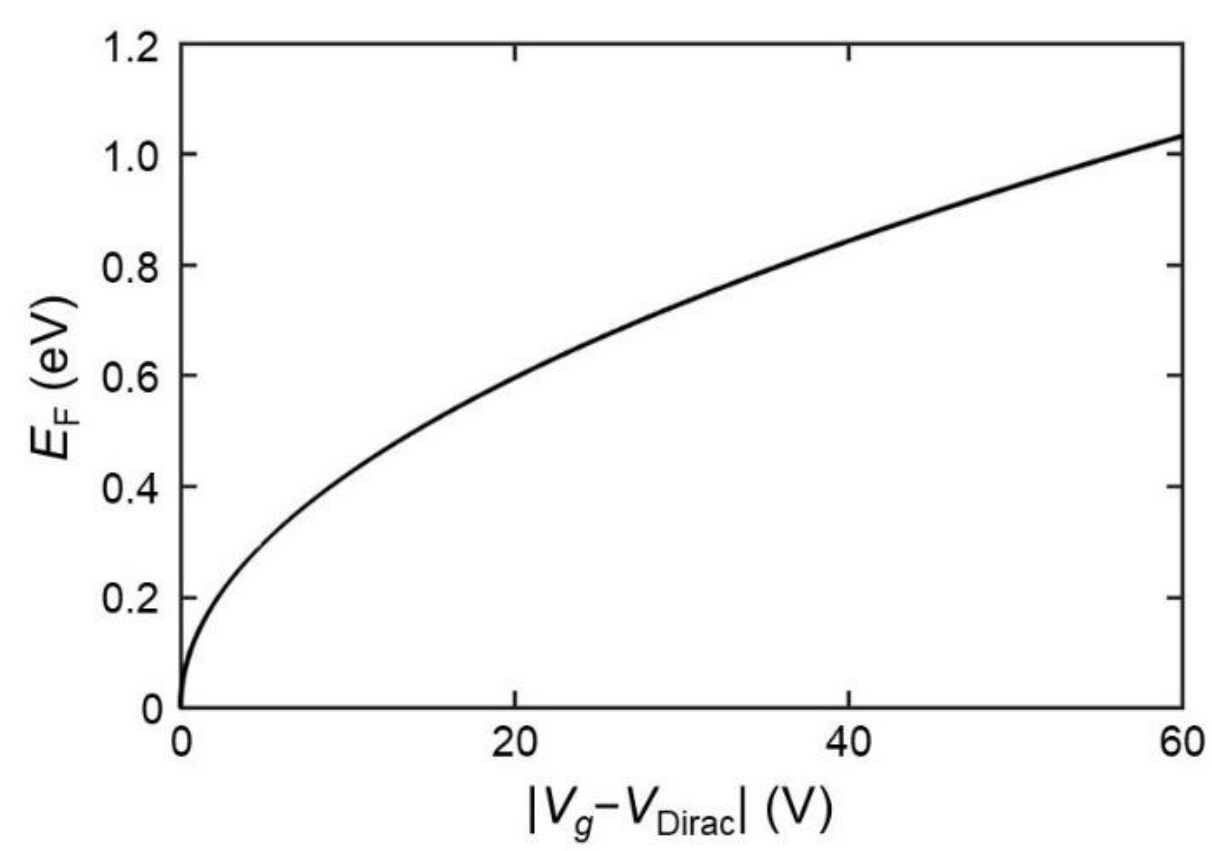

**Figure S1.** Dependence of graphene Fermi level as a function of gate voltage for the graphene–$HfO_2$–graphene capacitor structure

The two-layer graphene capacitor configuration enhances the modulation efficiency compared with a single-layer graphene configuration. By employing graphene as both the top and bottom electrodes of the $HfO_2$ dielectric, the interaction with the resonant system is doubled relative to a single-layer configuration in which only the bottom graphene is gated, resulting in a larger resonance shift for the same Fermi-level variation (Figure S2a).

To validate the parallel-plate approximation, we perform electrostatic simulations of the vertical electric field distribution under an applied bias. The field is highly localized within the capacitor region, and the magnitude at the center of the patterned graphene closely matches that of an ideal parallel-plate capacitor, confirming the validity of the model (Figure S2b). Based on this validation, in all optical simulations, the Fermi level is uniformly assigned across the patterned graphene region, while the unpatterned region (where the top graphene has been removed) is assumed to remain intrinsic ($E_F$ = CNP).

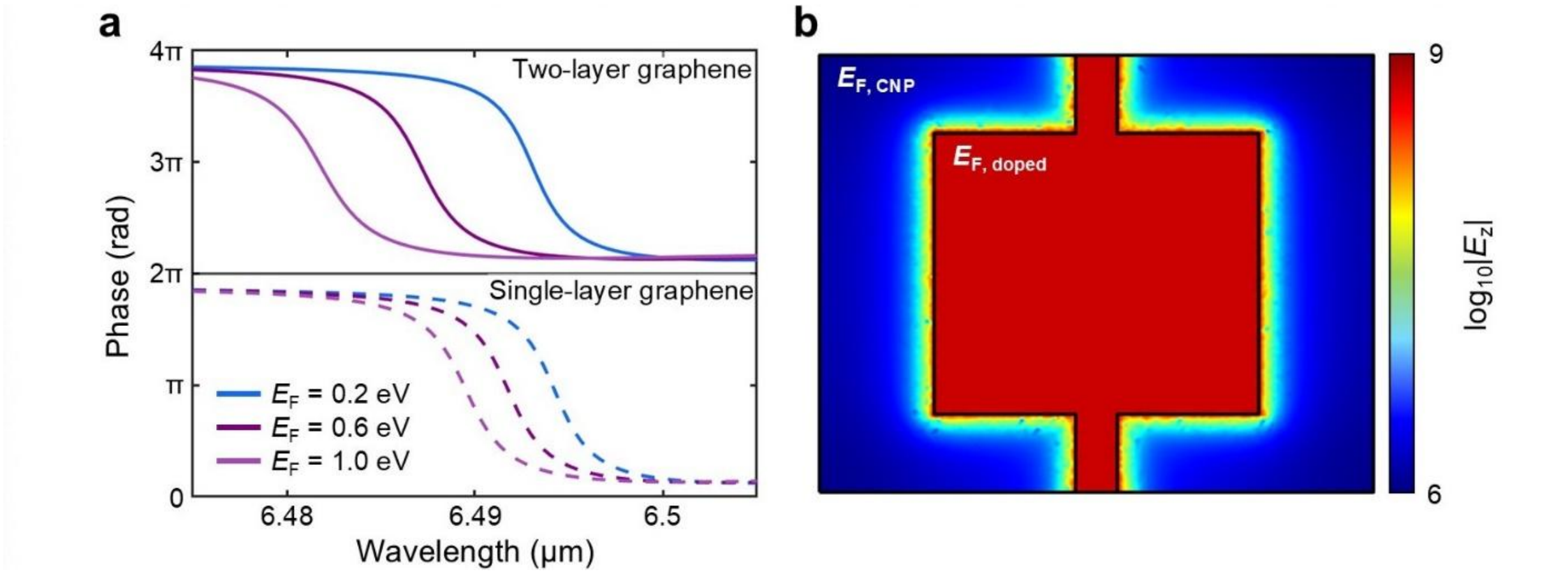

**Figure S2.** Graphene capacitor architecture and electrostatic field localization (a) Simulated phase response between of two-layer graphene capacitor configuration (solid lines) compared with the single-layer graphene configuration (dashed lines) at different Fermi levels ($E_F$ = 0.2, 0.6, and 1.0 eV). (b) Electrostatic simulation of the vertical electric field distribution (log10$|E_z|$) plotted on a z-cut plane at the center of the $HfO_2$ layer, under a 50 V bias applied to the top graphene layer.

**S2. Six-port single-resonance system**

To describe the complex transmission characteristics and Fano resonance of the system, we employ the temporal coupled-mode theory (TCMT)[2]. While the theoretical framework presented in the main text is based on a four-port single-resonance system, the optimized metasurface operates as a six-port single-resonance system. The physical principles and constraints, however, remain identical. Ports 1 and 2 correspond to the zeroth-order reflection and transmission, respectively, while ports 3–6 correspond to the first-order diffraction channels on the reflection side. Owing to the parity symmetry along the $x$- and $y$-directions, ports 3 and 4 share the same diffraction angle, as do ports 5 and 6 (Figure S3a–c). This symmetry also enforces equal radiative decay rates between ports 3 and 4, and between ports 5 and 6.

To establish the baseline performance of the lossless multi-port single-resonance system, the transmission characteristics of the optimized structure without graphene are shown in Figure S3d, e. In this configuration, the complex transmission amplitude circle shows a near-$2\pi$ phase coverage with a sharp resonance peak (Figure S3d, e). This confirms that the multi-port single-resonance system itself provides a high-Q resonance with a complex transmission amplitude circle that encircles the origin, establishing the foundation for dynamic independent amplitude and phase modulation.

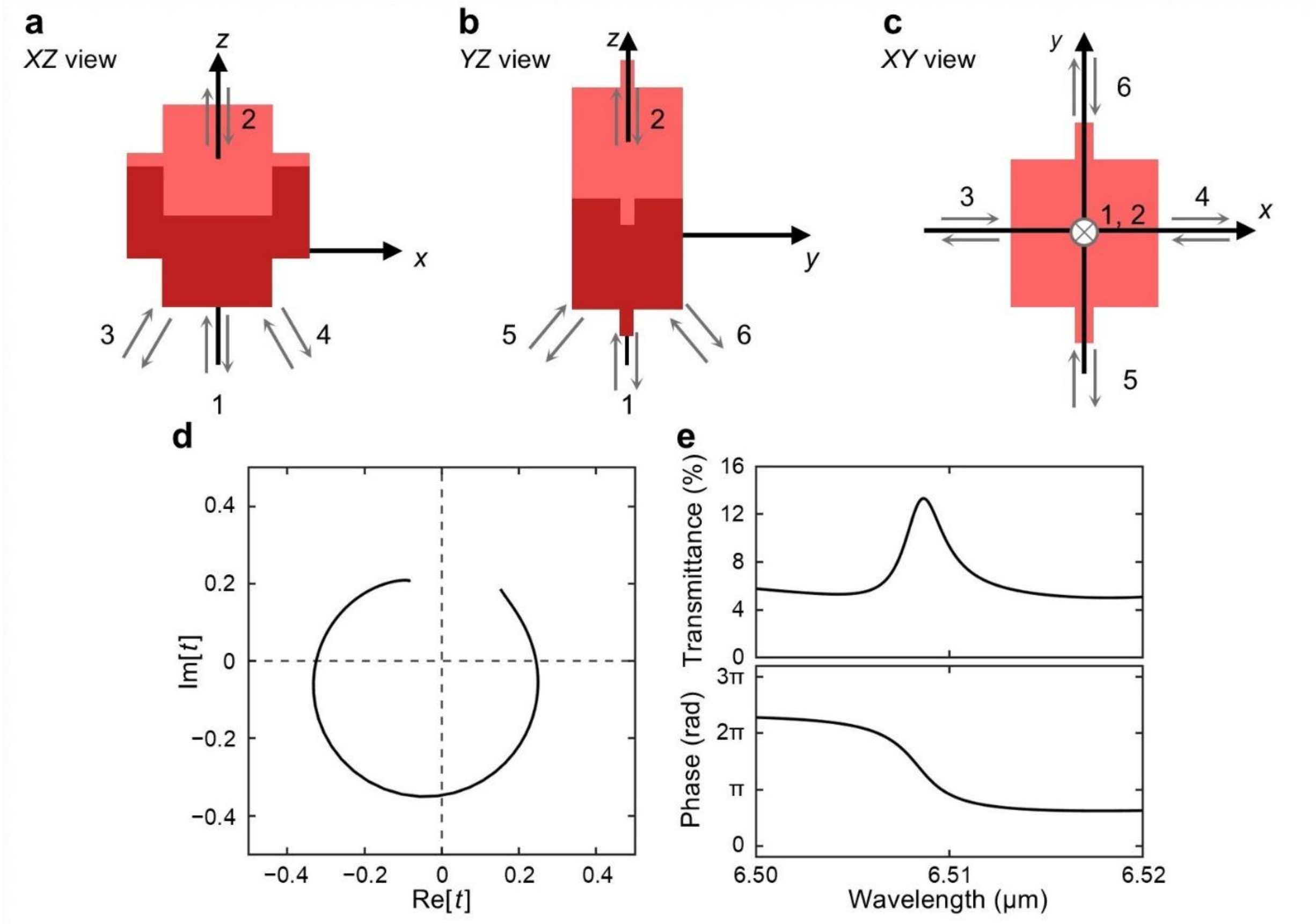

**Figure S3.** Six-port single-resonance system. (a–c) Schematic illustrations of the port configuration in the (a) *X–Z* view, (b) *Y–Z* view, and (c) top view, identifying the six radiative channels of the multi-port system. (d) Complex transmission amplitude circle in the complex plane for the optimized structure without graphene, showing near-2π phase coverage. (e) Corresponding transmittance (top) and phase (bottom) spectra, exhibiting a sharp Fano-like resonance characteristic of the high-Q resonance system.

To rigorously characterize the resonance dynamics upon graphene integration, the TCMT parameters are extracted based on the temporal coupled-mode theory (TCMT) framework. First, the total decay rate and resonance frequency at each Fermi level are extracted from the Fano formula[3] (Figure S4a). The individual radiative decay rates are obtained by fitting the complex transmission amplitude circle in the complex plane of the reflection and transmission coefficient associated with each port. For each radiative decay rate, the radius of the corresponding circle is given by $R = \sqrt{\gamma_{\text{in}}\gamma_{\text{out}}}/\gamma$, where $\gamma_{\text{in}}$ and $\gamma_{\text{out}}$ denote the radiative decay rate into the input and output ports, respectively.[4] The dissipative decay rate $\gamma_d$ is then determined by subtracting the sum of the radiative decay rates from the total decay rate obtained from the Fano fitting (Figure S4b).

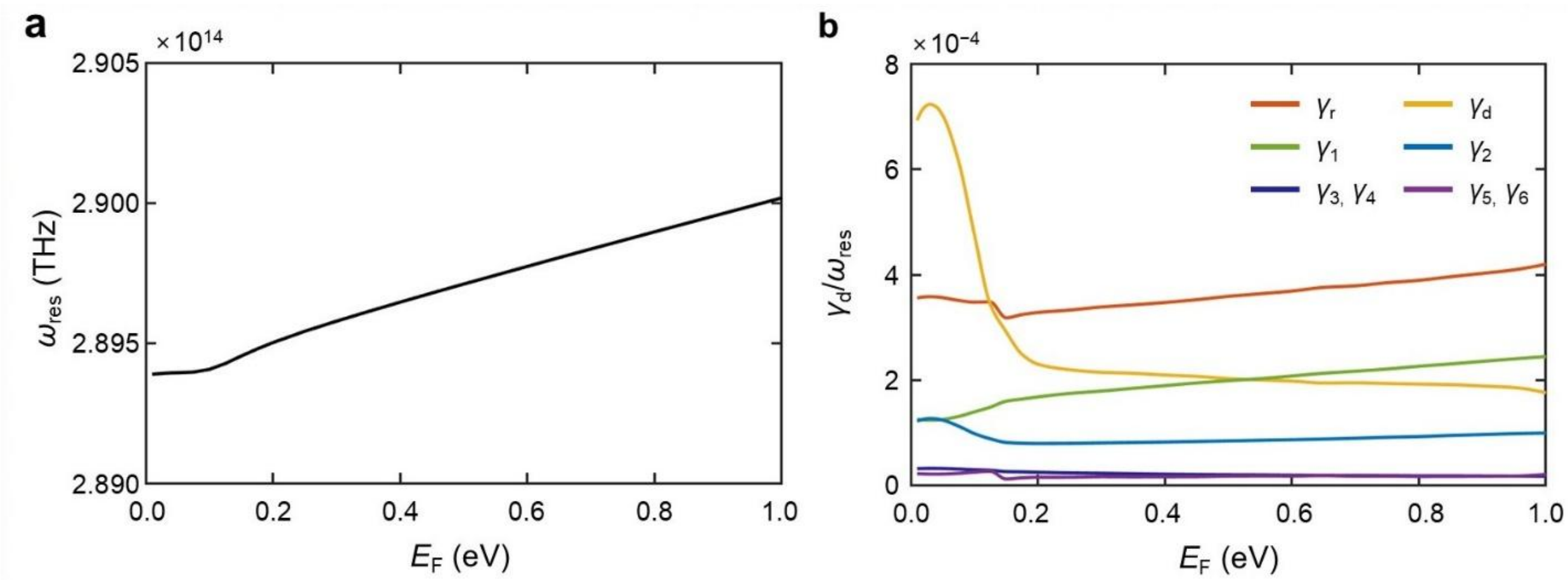

**Figure S4**. Extracted TCMT parameters as a function of graphene Fermi level. (a) Resonance frequency ($\omega_{res}$) as a function of graphene Fermi level ($E_F$). (b) Radiative decay rates ($\gamma_1$–$\gamma_6$) and dissipative decay rate ($\gamma_d$) as a function of graphene Fermi level ($E_F$).

The extracted dissipative decay rate exhibits the same trend as the real part of graphene's optical conductivity, confirming that $\gamma_d$ is governed by Re($\sigma$) (Figure S5a). Likewise, the resonance frequency shift ($\Delta\omega_{res}$) follows the same trend as the imaginary part of graphene's optical conductivity (Im($\sigma$)) (Figure S5b) confirming that $\omega_{res}$ is governed by Im($\sigma$). Consequently, the low-Fermi-level regime enables modulation of the dissipative decay rate, while the high-Fermi-level regime enables modulation of the resonance frequency.

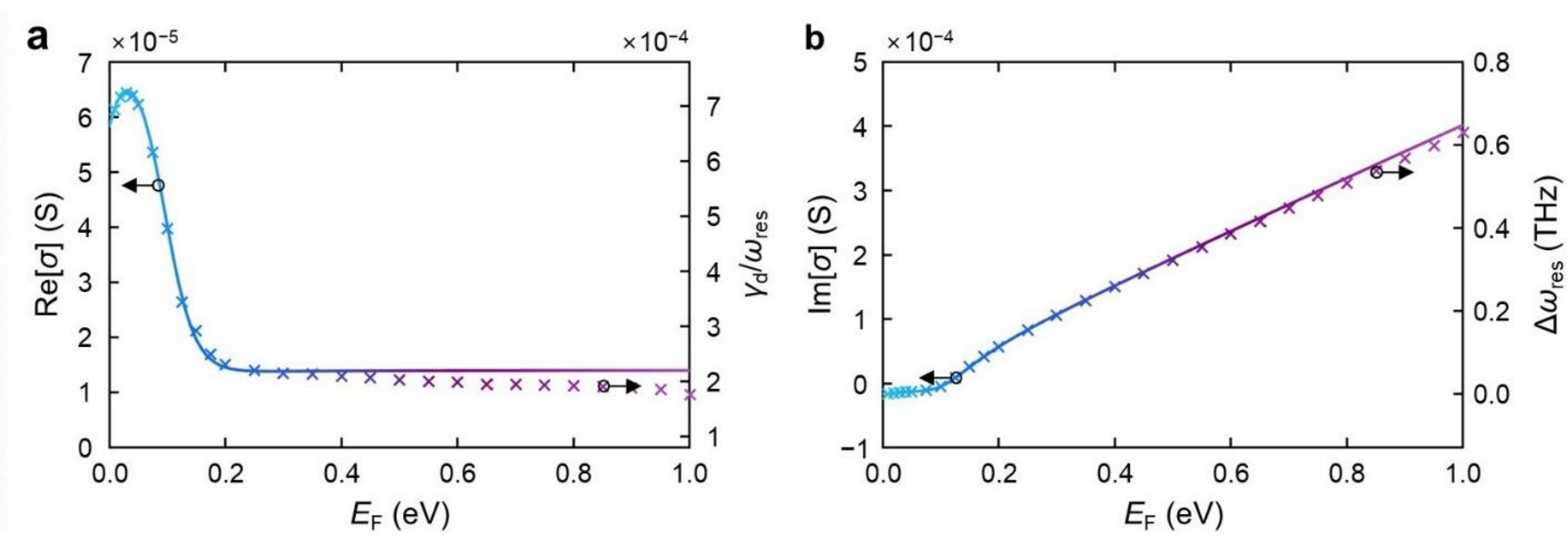

**Figure S5.** Correlation between graphene conductivity and TCMT parameters. (a) Real part of graphene's optical conductivity (Re($\sigma$)) and the extracted dissipative decay rate ($\gamma_d$) as a function of Fermi level. (b) Imaginary part of graphene's optical conductivity (Im($\sigma$)) and the resonance frequency shift ($\Delta\omega_{res}$) as a function of Fermi level.

The quality factor (Q-factor) can also be extracted from the Fano formula as $Q = \omega_{res}/\gamma_0$. The extracted Q-factor increases significantly as the system transitions from the high-loss interband regime to the low-loss intraband regime, ranging from 800 to 1,600 before stabilizing between 1,600 and 1,800 at high Fermi levels (Figure S7).

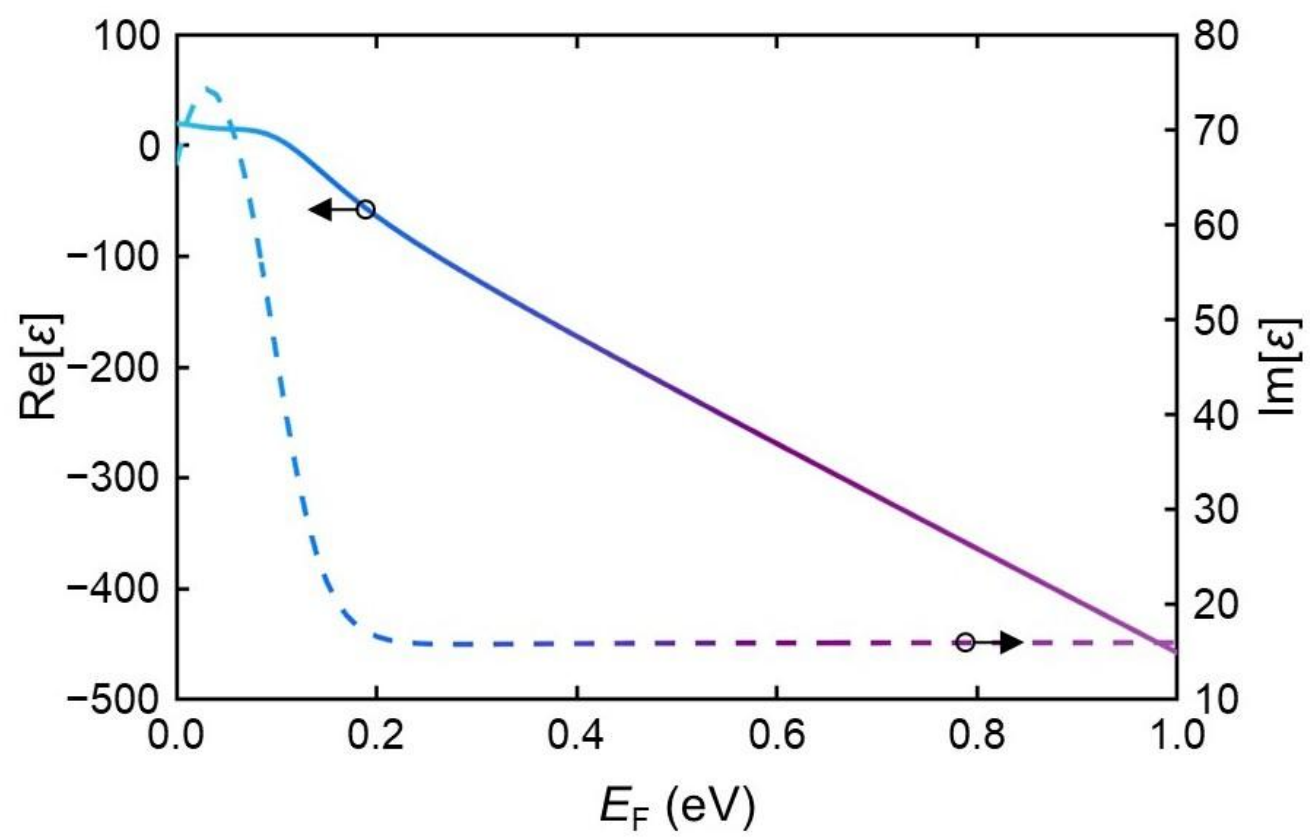

**Figure S6.** Real (solid line, left axis) and imaginary (dashed line, right axis) parts of graphene's effective permittivity as a function of Fermi level ($E_F$) at $\lambda = 6.5$ µm, calculated from the optical conductivity via $\varepsilon = \varepsilon_0 + i\sigma/(\omega t_g)$. Re($\varepsilon$) follows the trend of Im($\sigma$) and governs the effective refractive index, thereby controlling the resonance frequency $\omega_{res}$, while Im($\varepsilon$) follows the trend of Re($\sigma$) and governs the absorption, thereby controlling the dissipative decay rate $\gamma_d$.

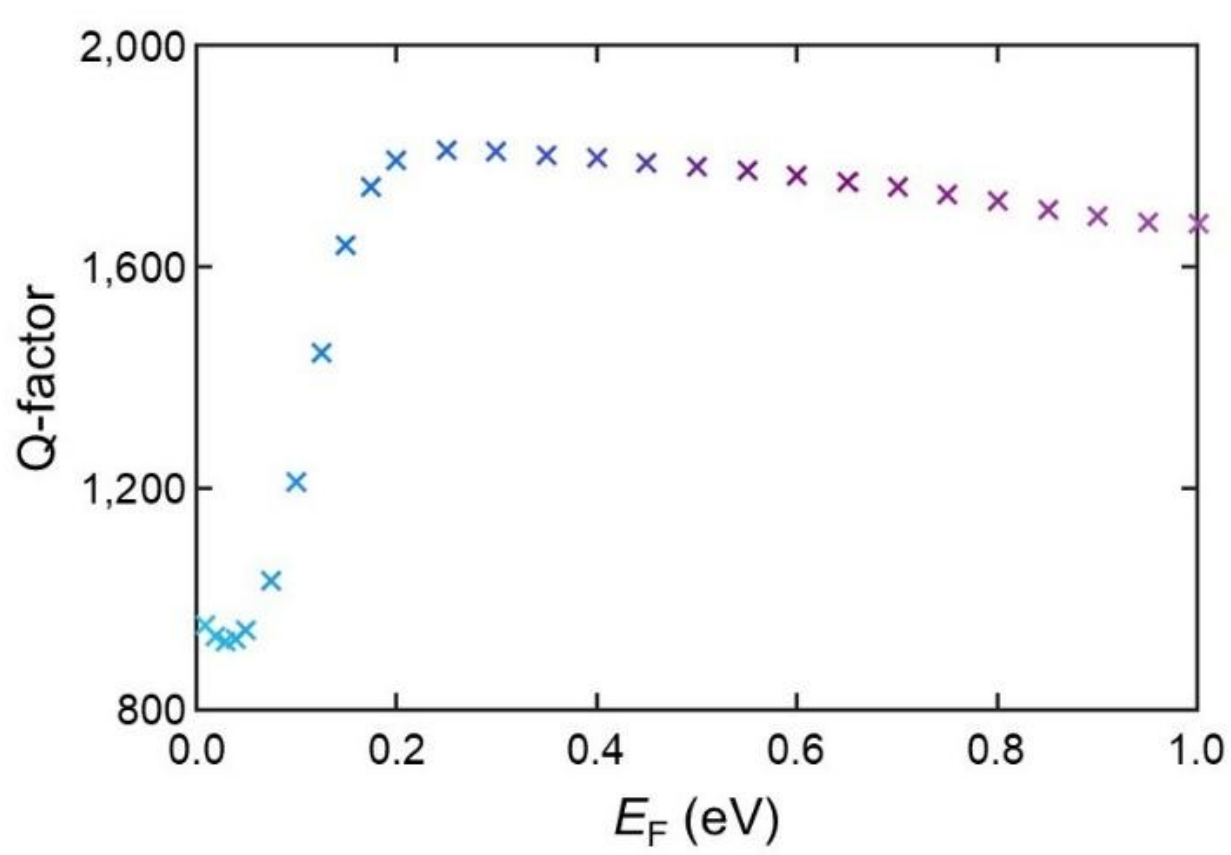

**Figure S7.** Q-factor of the resonance as a function of graphene Fermi level.

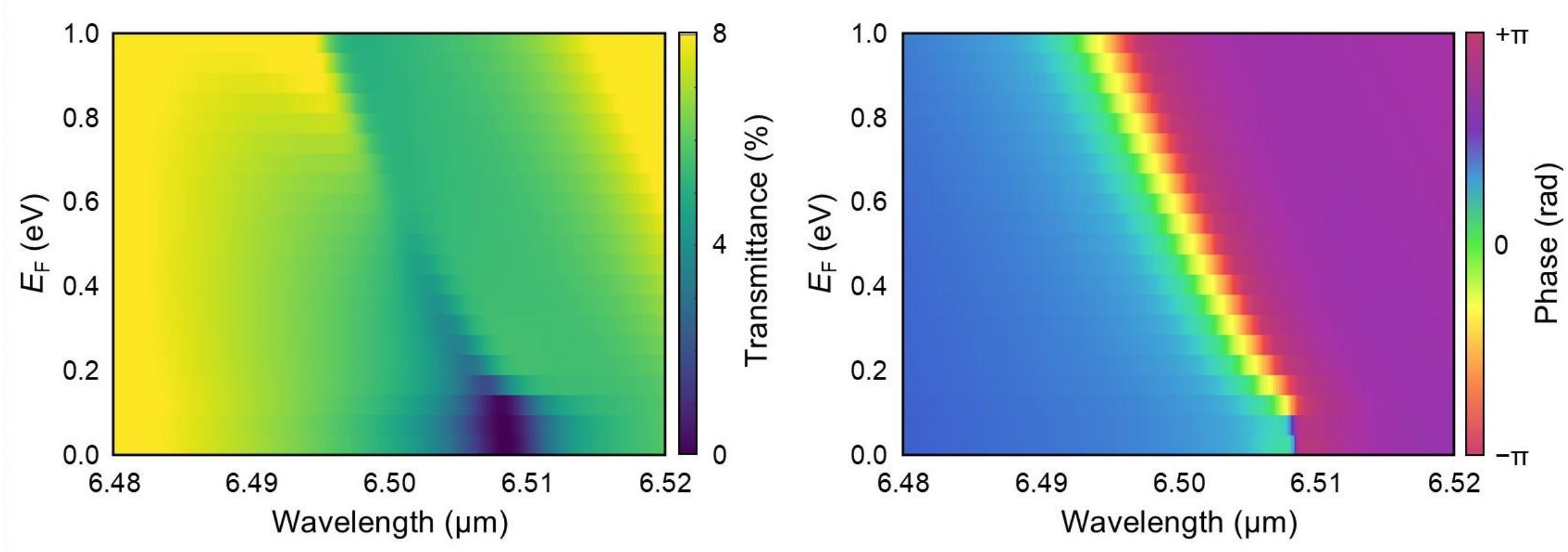

**Figure. S8**. Transmittance (left) and phase (right) of the optimized structure as a function of wavelength and graphene Fermi level ($E_F$). At high Fermi levels, the low dissipative loss keeps the complex transmission amplitude circle large enough to encircle the origin, yielding a continuous 2π phase coverage across the resonance. As the Fermi level decreases, the increasing dissipative decay rate contracts the circle, until the trajectory eventually passes through the origin, producing a transmission null accompanied by an abrupt π phase jump (dark region in the left panel). Below this point, the resonance circle no longer encloses the origin, and the spectral phase coverage collapses to less than π.

## S3. Dependence of amplitude and phase modulation with different graphene mobility

To demonstrate the viability of the metasurface across experimentally accessible graphene qualities, we assess the influence of carrier mobility on the modulation performance. Figure S9 shows the complex transmission coefficient trajectories as the graphene Fermi level is varied at the two operating wavelengths for amplitude modulation ($\lambda$ = 6.509 μm) and phase-only modulation ($\lambda$ = 6.5 μm), using the optimized geometric parameters and carrier mobility ($\mu_s$ = 1,000 $cm^2\ V^{-1}\ s^{-1}$) from Figure 3 of the main text. Since the modulation performance is governed by graphene's optical conductivity, we first examine how the conductivity varies with carrier mobility.

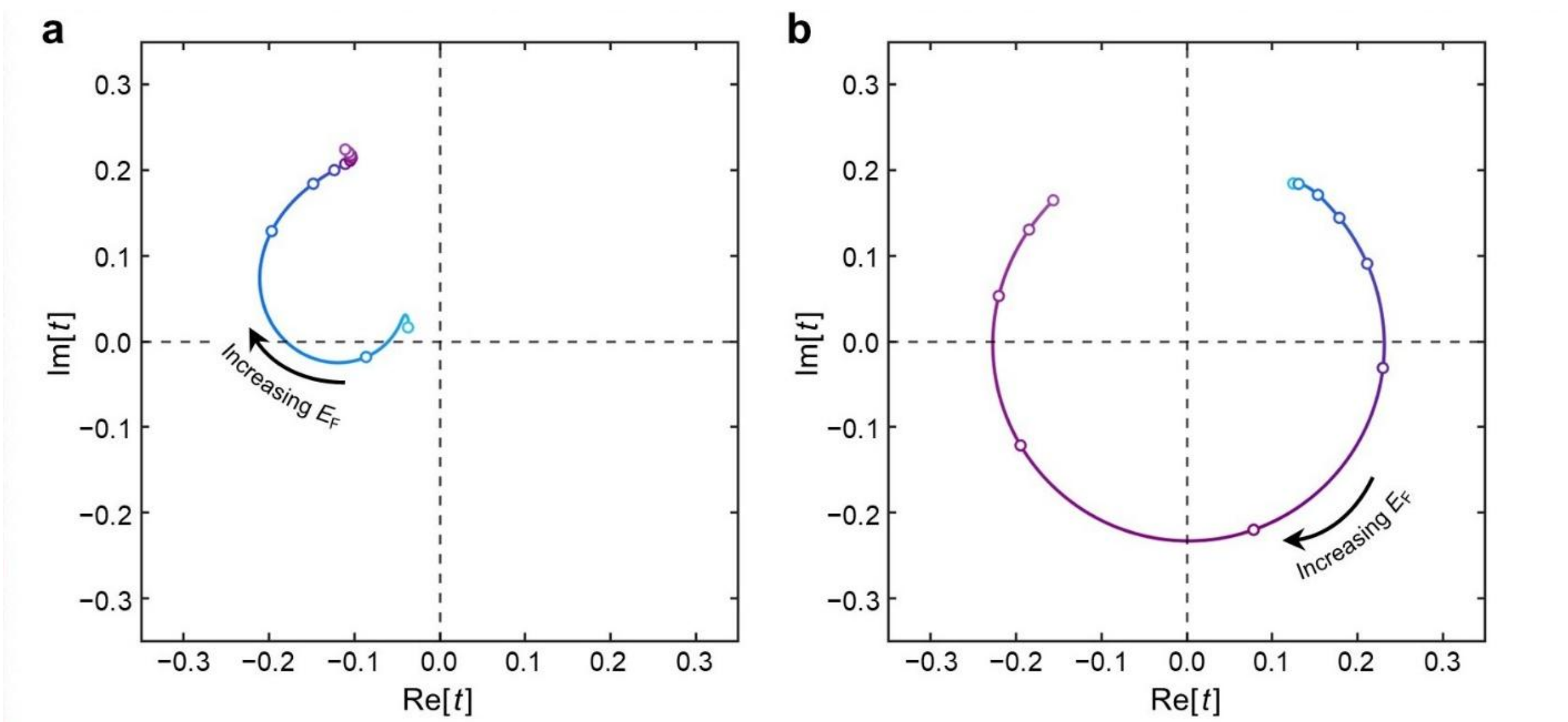

**Figure S9.** Complex transmission coefficient trajectories as the graphene Fermi level increases. (a) Trajectory at $\lambda$ = 6.509 μm, corresponding to the amplitude modulation wavelength. The inset shows a magnified view near the origin, where the trajectory passes through zero, enabling transmission modulation efficiency of ~100%. (b) Trajectory at $\lambda$ = 6.5 μm, corresponding to the phase-only modulation wavelength, demonstrating near-circular coverage for phase-only modulation. Open circles represent discrete $E_F$ steps of 0.1 eV.

As shown in Figure S10a, b, the real part of graphene's conductivity, Re($\sigma$), remains similar across different mobilities in the low-Fermi-level regime, but decreases with increasing mobility in the high-Fermi-level regime, while the imaginary part, Im($\sigma$), remains nearly invariant across the considered mobility range. In particular, at high Fermi levels where intraband transitions dominate, higher mobility reduces the scattering rate and thus Re($\sigma$), resulting in lower dissipative decay rate. Consequently, lower mobility increases the system loss, contracting the radius of the complex transmission amplitude circle, whereas the resonance peak position remains essentially unchanged.

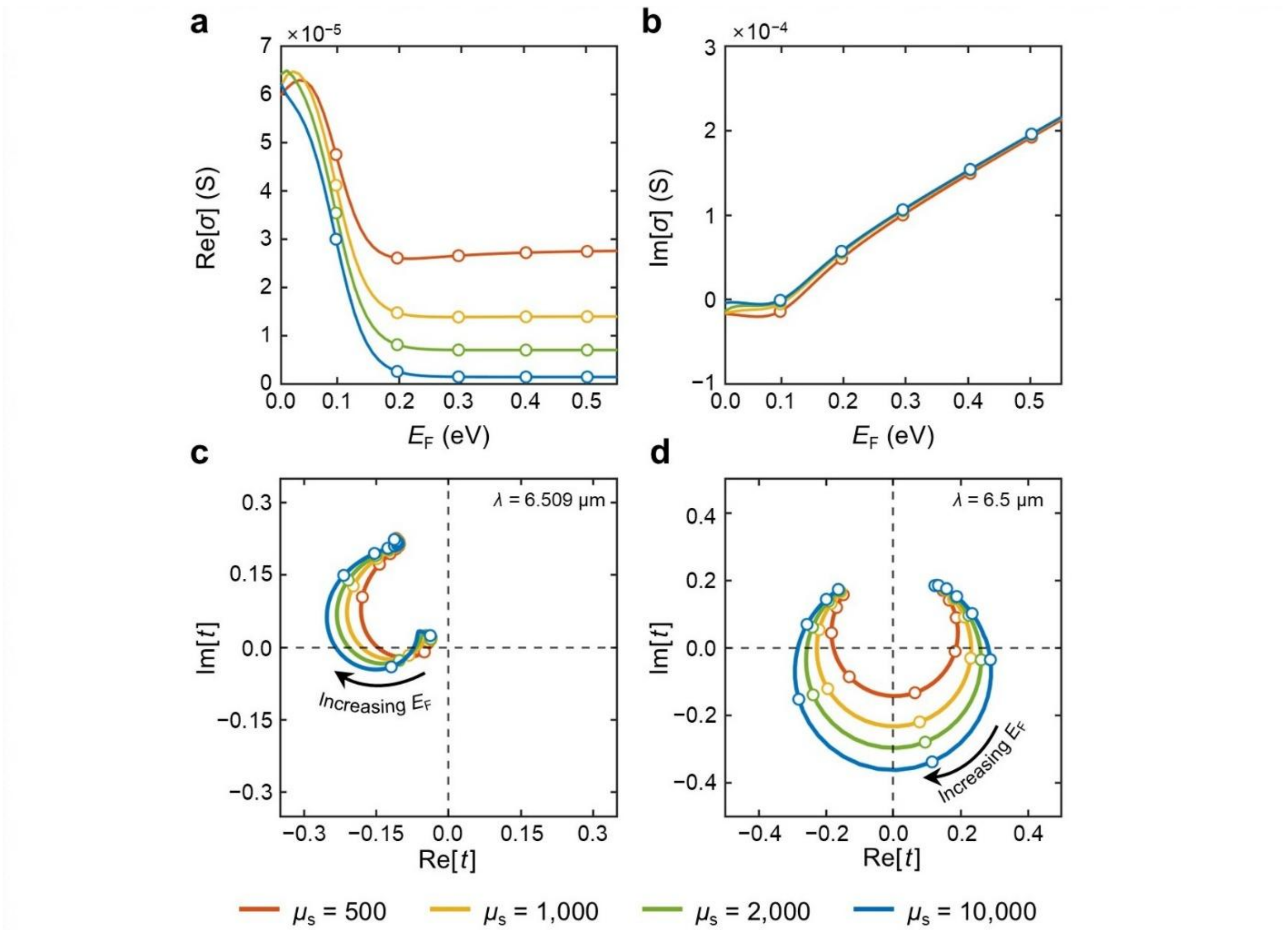

**Figure S10.** Dependence of graphene conductivity and complex transmission trajectory on graphene carrier mobility. (a) Real and (b) imaginary parts of graphene conductivity as a function of Fermi level ($E_F$) at $\lambda$ = 6.5 μm for different mobilities. (c) Complex transmission coefficient trajectories at $\lambda$ = 6.509 μm, corresponding to the amplitude modulation wavelength, and (d) at $\lambda$ = 6.5 μm, corresponding to the phase modulation wavelength. The structure uses the same geometric parameters as in Figure 3 of the main text, with the graphene Fermi level varied for different carrier mobilities ($\mu_s$ = 500, 1,000, 2,000, and 10,000 cm$^2$ V$^{-1}$ s$^{-1}$). Open circles represent discrete $E_F$ steps of 0.1 eV.

For amplitude modulation, the loss difference directly governs the radius of the resonance trajectory and thus determines the maximum achievable transmittance. Specifically, higher graphene mobility suppresses the intraband transitions, resulting in a monotonic decrease in the real part of the conductivity even within the $E_F$ range from the CNP to 0.1 eV. Driven by this loss reduction, the transmission coefficient trajectory follows a clearer radial path outward in the complex plane. This enables a higher maximum transmittance, demonstrating that higher mobility directly enhances the modulation efficiency of amplitude modulation. (Figure S10c).

For phase modulation, the resonance shift remains similar across different mobilities, indicating that the achievable phase modulation range is less dependent on mobility. However, the reduced loss at higher mobility enlarges the resonance circle, introducing amplitude variations along the trajectory. This variation can be eliminated by optimizing the geometric parameters to re-center the transmission circle, resulting in a increased transmittance (Figure S10d).

## S4. Figure of merit of phase-only modulation

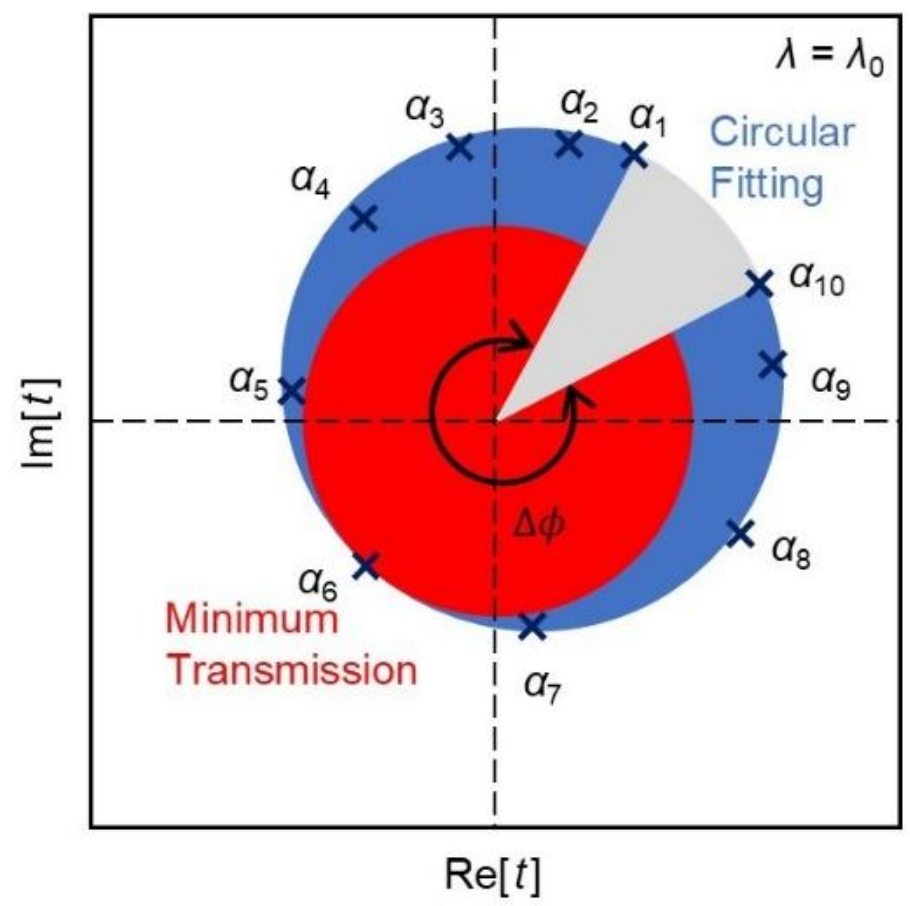

**Figure S11.** Figure of merit for phase-only modulation. Schematic illustration of the FoM definition in the complex transmission amplitude plane, where $\alpha$ denotes the tuning parameter (graphene Fermi level or temperature) that varies along the trajectory. The minimum transmission point and the circular fitting of the trajectory are indicated.

The figure of merit (FoM) is used to optimize the geometric parameters of the metasurface for phase-only modulation performance. Although two distinct tuning mechanisms are presented in the main text, the underlying definition of the FoM remains invariant across different tuning mechanisms. The primary objective is to maximize the phase modulation range while maintaining high and constant transmittance. To mitigate the computational burden of simulating a continuous sweep, the FoM is evaluated based on transmission coefficients sampled at 10 (Fermi level optimization) or 11 (temperature optimization) linearly spaced discrete points within the tuning range.

We adopt a geometric approach in the complex plane that simultaneously captures both the phase coverage and the transmittance of the modulation trajectory. As the tuning parameter is varied, the transmission coefficient traces a trajectory in the complex plane, and the area swept by this trajectory with respect to the origin ($A_{\mathrm{swept}}$) encodes both the phase modulation range and the transmittance at each tuning step. This swept area is given by

$$A_{\mathrm{swept}} = \frac{1}{2}\int |t(\phi)|^2 d\phi, \quad (4)$$

which represents the exact sector area enclosed by connecting the origin to every point along the trajectory. A larger swept area corresponds to either a wider phase coverage, a higher transmittance, or both, making it a natural quantity to maximize for phase-only modulation.

However, directly evaluating this integral requires a continuous sampling of the trajectory, which is computationally expensive. To circumvent this limitation, we approximate the swept area by leveraging the near-circular morphology of the trajectory, which allows the area to be computed analytically from the geometric parameters of a fitted circle. Although the trajectory of the complex transmission coefficient as a function of the tuning parameter is not strictly circular, the complex transmission amplitude trajectory in the frequency domain is circular. As the phase modulation mechanism relies on peak shifting with a relatively stable decay rate, the resulting locus in the complex plane exhibits a near-circular morphology, justifying the validity of circle fitting. Furthermore, this fitting enables the interpolation of intermediate states, providing a continuous approximation of the device's performance from a limited set of discrete simulations.

The discrete transmission coefficients obtained for each tuning step are fitted to a circle described by $(x - a)^2 + (y - b)^2 = r^2$, where $(a, b)$ is the center of the fitted circle and $r$ is the fitted radius. The swept area is then obtained by subtracting the unswept region of the fitted circle (gray area in Figure S11) from the total fitted circle area, which gives

$$A_{\mathrm{swept}} \approx \pi r^2 - A_{\mathrm{gray}} \quad (5)$$

where $A_{\mathrm{gray}}$ corresponds to the portion of the fitted circle that lies outside the modulation range $\Delta\varphi$.

Within this framework, $A_{\mathrm{swept}}$ is decomposed into two components, the red area ($A_{\mathrm{red}}$) and the blue area ($A_{\mathrm{blue}}$), as illustrated in Figure S11. The red area represents the guaranteed baseline transmittance and is defined as the inner circular sector determined by the minimum transmission amplitude over the modulation range, and is given by

$$A_{\mathrm{red}} = \frac{1}{2}|t_{\mathrm{min}}|^2 \times \Delta\phi \quad (6)$$

where $|t|_{\mathrm{min}}$ is the minimum transmission amplitude and $\Delta\varphi$ is the phase modulation range. Ideally, this baseline area should be maximized to ensure a high transmittance across the modulation range. The blue area represents the undesirable amplitude variation above this baseline, and follows as

$$A_{\mathrm{blue}} = A_{\mathrm{total}} - A_{\mathrm{red}}, \quad (7)$$

which captures the integrated amplitude fluctuation along the trajectory. Minimizing $A_{\mathrm{blue}}$ ensures that the complex transmission coefficient trajectory remains close to the minimum baseline circle, thereby suppressing amplitude ripples and maintaining a constant transmittance.

Finally, the FoM is constructed to reward a large baseline area while penalizing the variation area, weighted by the square of the phase modulation, and is defined as

$$FoM = (A_{\mathrm{red}} - A_{\mathrm{blue}}) \times \Delta\phi^2 \quad (8)$$

where $\Delta\varphi$ is squared to prioritize the maximization of the phase shift. By maximizing this area-based FoM, the optimization process yields a trajectory that achieves a large phase modulation with high and constant transmittance, effectively balancing the trade-off between phase modulation and amplitude stability. It should be noted that this FoM is strictly applied within the parameter space where the complex transmission trajectory encircles the origin, starting from an initial geometric design whose complex transmission trajectory encircles the origin.

## S5. Independent control of resonance frequency and dissipative decay rate via dual tuning

In the main text, graphene electrostatic gating serves as the tuning mechanism for both amplitude and phase modulation, with Ge thermo-optic tuning employed as a global tuning mechanism that spectrally aligns the two operating wavelengths. Here we examine an alternative configuration in which each TCMT parameter is assigned to a distinct tuning mechanism. Although graphene enables $\omega_{res}$ tuning in the high-Fermi-level regime, the accompanying variation in the radiative decay rate alters the size of the complex transmission amplitude circle, deviating from ideal phase-only modulation. In contrast, Ge thermo-optic tuning exclusively controls $\omega_{res}$ without perturbing the decay rates, making thermo-optic tuning ideally suited for phase-only modulation.

To verify this configuration, we re-optimize the geometric parameters with Ge thermo-optic tuning as the phase-only modulation mechanism. The optimized device confirms that the thermo-optic effect modulates only the resonance frequency, leaving the radiative and dissipative decay rates unperturbed (Figure S12a). This stability extends across a finite spectral window (6.505 μm < λ < 6.512 μm), within which the device maintains a phase modulation exceeding 280° with transmittancevariation below 15% under a temperature change of only 25 K (Figure S12b), highlighting the robustness of thermo-optic tuning as an ideal $\omega_{res}$ tuning mechanism. Conversely, lowering the graphene Fermi level at a fixed temperature contracts the radius of the complex transmission amplitude circle, eventually driving the system to a transmission null (Figure S12c). The corresponding spectral response confirms that the transmission null is achieved at low Fermi levels (Figure S12d), demonstrating that graphene functions as a loss-control layer.

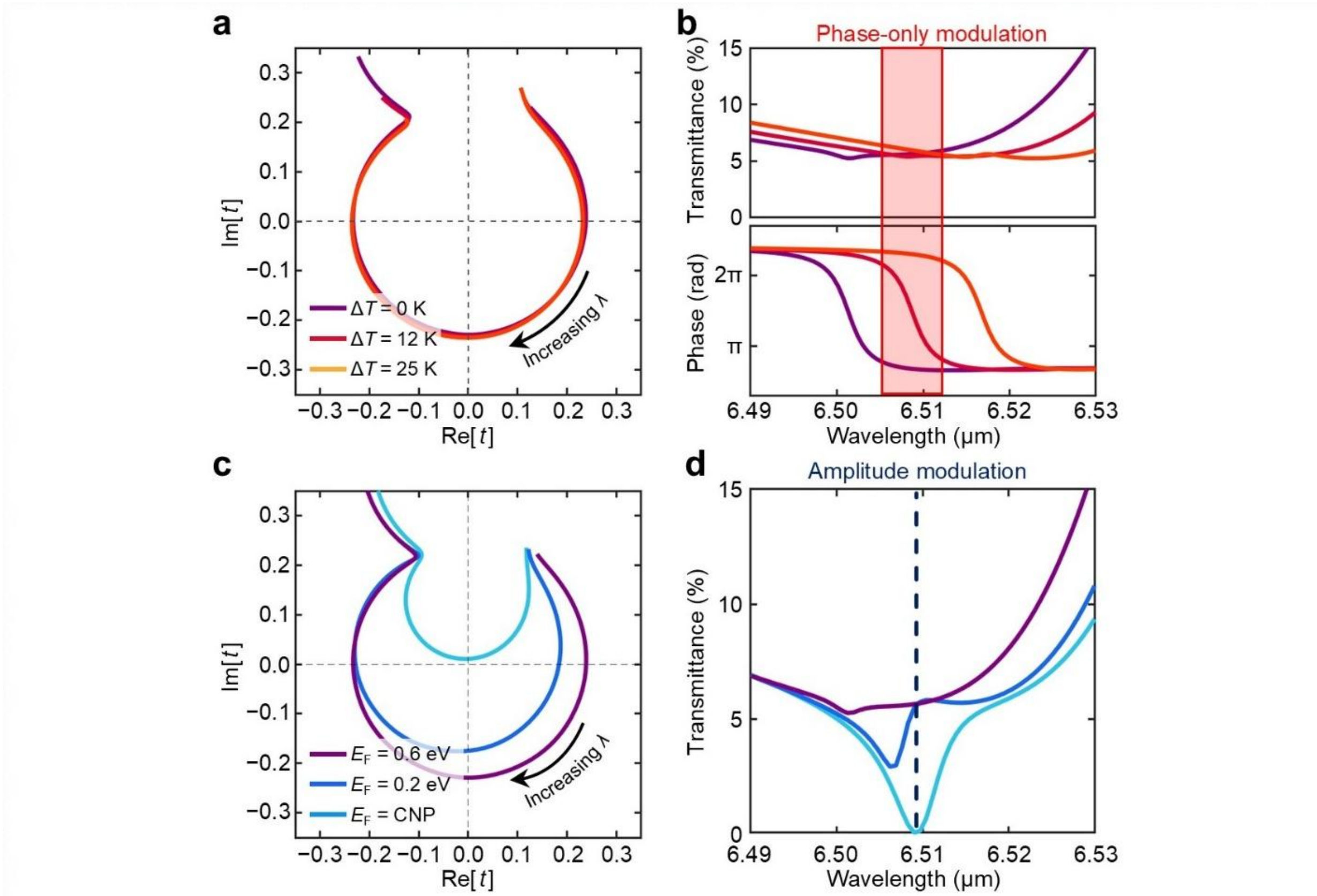

**Figure S12.** Distinct TCMT parameter control via physically different tuning mechanisms (a) Complex transmission coefficient circles in the complex plane for varying temperature at $E_F = 0.6$ eV. (b) Spectral response of transmittance (top) and phase (bottom) for the corresponding temperature variations showing stable decay rate with only $\omega_{res}$ tuning allowing broad phase-only modulation. (c) Complex transmission coefficient circles in the complex plane for varying graphene Fermi level at $T = 293$ K. (d) Spectral response of transmittance showing the transmission null at low Fermi level. The geometric parameters are $P_x = 6.438$ μm, $P_y = 5.844$ μm, $w = 3.76$ μm, $l = 3.754$ μm, $h = 3.38$ μm, $w_c = 0.482$ μm, $t = 0.087$ μm.

The integrated operation of the two mechanisms is mapped across the temperature–Fermi level parameter space at $\lambda = 6.51$ μm (Figure S13). Along Path 1 ($E_F = 0.6$ eV, varying temperature), the thermo-optic effect yields a phase modulation of 288° with the transmittance maintained at 5% and transmittance variation below 8%. Along Path 2 ($T = 294.6$ K, varying $E_F$), lowering the Fermi level increases the dissipative decay rate, contracting the resonance trajectory and driving the transmittance from 6% to a near-zero state ($T_{min} = 0.002\%$, $T_{max} = 6.2\%$), effectively crossing the origin. Path 3 ($E_F = 0.07$ eV, varying temperature) realizes amplitude modulation via $\omega_{res}$ tuning at a reduced Fermi level. In this path, the increased dissipative decay rate at the lower Fermi level drives the system into a regime where the resonance exhibits a transmission dip, and subsequent $\omega_{res}$ tuning via temperature variation sweeps the transmission through zero, thereby realizing a transmittance modulation efficiency of 100% ($T_{min}$ = 0.001%, $T_{max}$ = 5.7%). Unlike Path 2, this route achieves amplitude modulation through $\omega_{res}$ tuning rather than $\gamma_d$ tuning so the accompanying resonance shift couples the phase to the amplitude response. Collectively, these three modulation paths demonstrate that the dual-tuning framework supports multiple modulation routes at a single operating frequency—phase-only modulation via $\omega_{res}$ tuning (Path 1), amplitude modulation via $\gamma_d$ tuning (Path 2), and amplitude modulation via $\omega_{res}$ tuning (Path 3). More broadly, these routes are not tied to a specific tuning mechanism, and any pair of mechanisms that independently access $\omega_{res}$ and $\gamma_d$ can realize independent amplitude and phase modulation.

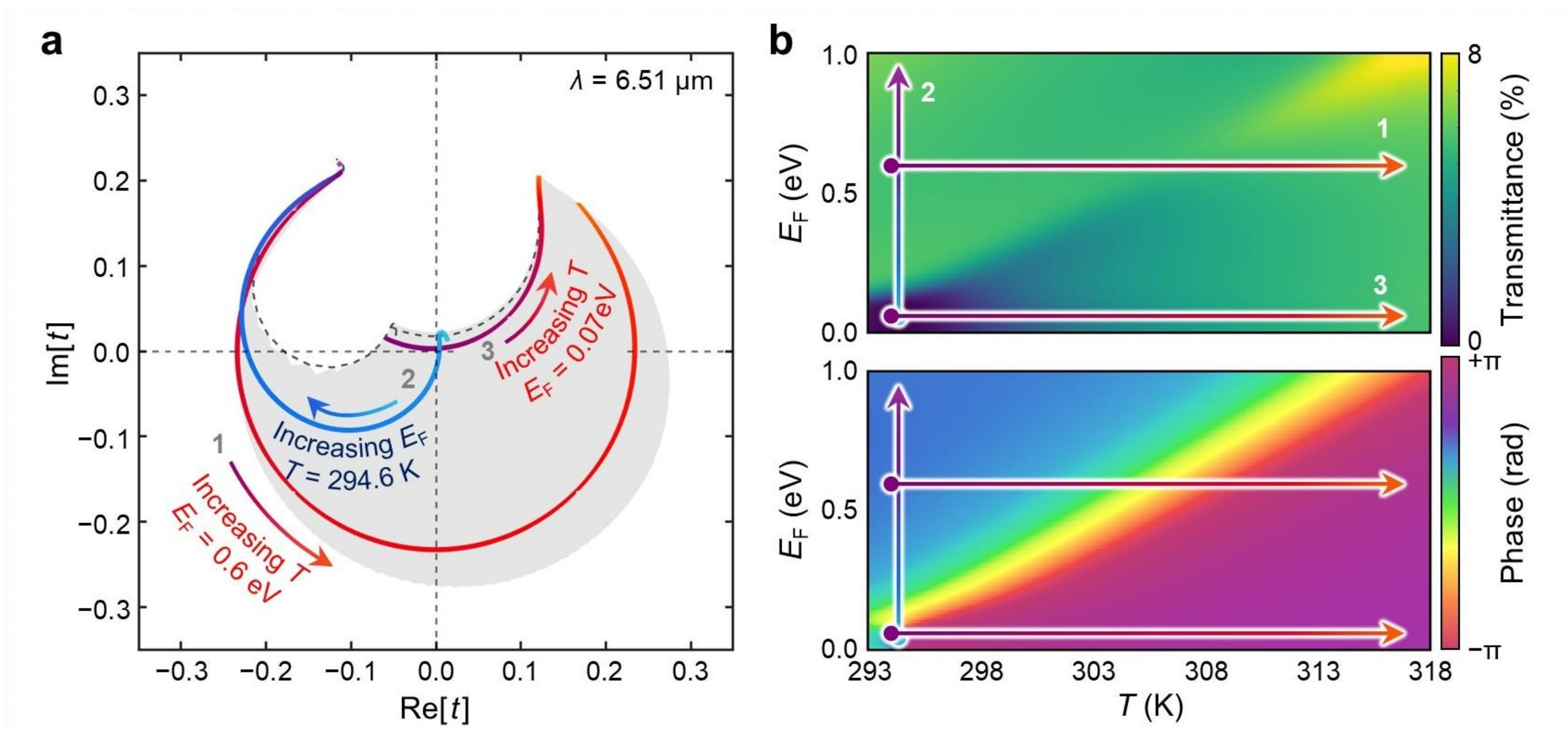

**Figure S13.** Phase-only and amplitude modulation via dual tuning mechanisms at a fixed wavelength. (a) Complex transmission coefficient trajectories in the complex plane at $\lambda = 6.51$ µm, showing three modulation paths in the temperature–Fermi level parameter space. Path 1 ($E_F = 0.6$ eV, varying temperature) demonstrates phase-only modulation with nearly 290° phase modulation while maintaining a stable transmittance of 5% with transmittance variation below 8%. Path 2 ($T$ = 294.6 K, varying $E_F$) achieves amplitude modulation with transmittance modulation efficiency of ~100% via $\gamma_d$ tuning, driving the transmittance from 5.5% to near-zero. Path 3 ($E_F = 0.07$ eV, varying temperature) realizes amplitude modulation via $\omega_0$ tuning at low Fermi level. The shaded grey area indicates the accessible region in the complex plane spanned by both control parameters, confirming complete complex amplitude modulation. (b) Transmittance (top) and phase (bottom) mapped across the $T$–$E_F$ parameter space at $\lambda = 6.51$ µm, with the three modulation paths overlaid.

## S6. Two-level phase gradient metasurface

While the main text presents the gradient-descent optimization of the two-level phase gradient metasurface at the target wavelength $\lambda_1$ = 6.51 μm, here we examine the wavelength dependence of the unoptimized configuration. At $\lambda_1$, the two Fermi levels are selected to yield the ideal phase difference of π between adjacent unit cells (Figure S14a). However, the diffraction efficiency at this wavelength remains low due to near-field coupling between unit cells. Notably, a higher relative efficiency is observed at a slightly detuned wavelength $\lambda_2$ = 6.513 μm (Figure S14b), where the phase difference between the two Fermi levels is only 0.53π (Figure S14c). This counterintuitive result demonstrates that the ideal phase condition does not directly translate to high diffraction efficiency in the presence of near-field coupling, motivating the gradient-descent optimization at the target wavelength $\lambda_1$.

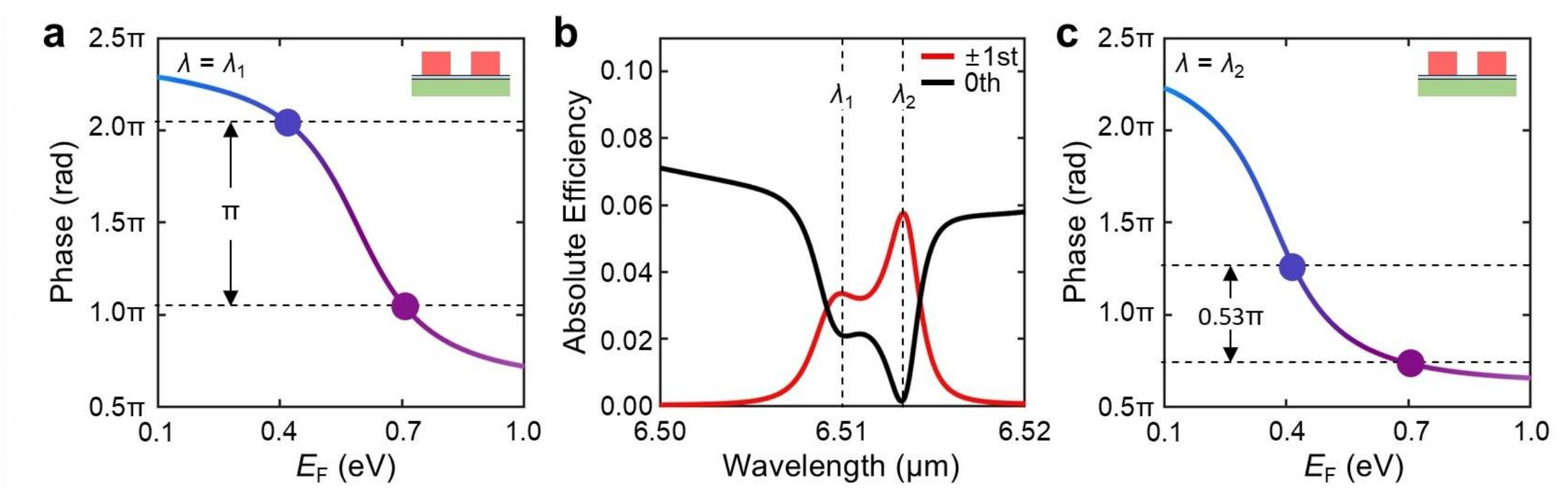

**Figure S14.** Initial condition for two-level phase grating without optimization. (a) Phase as a function of Fermi level at $\lambda = \lambda_1$, where the two selected Fermi levels (dots) yield a phase difference of π, corresponding to the ideal condition for a two-level phase grating. (b) Absolute efficiency of the zeroth-order (black) and ±1st-order (red) diffraction as a function of wavelength. At the target wavelength $\lambda_1$, the ±1st-order diffraction efficiency remains low despite the ideal π phase difference, due to near-field coupling between adjacent unit cells. A higher efficiency is observed at $\lambda_2$. (c) Phase as a function of Fermi level at $\lambda = \lambda_2$, where the phase difference between the two selected Fermi levels is 0.53π.

## S7. Relative efficiency of phase gradient metasurface

The phase gradient metasurface was optimized to maximize the relative efficiency of the first order diffraction. To optimize the relative efficiency, each unit cell's Fermi level was optimized, resulting 99.9%, 96.9%, and 93.1%, for two-, three-, and four-level phase gradient metasurfaces, respectively.

Figure S15 presents the relative efficiency of each diffraction order at the resonant wavelength for the optimized beam-steering metasurfaces shown in the main text. For all three quantization levels ($N$ = 2, 3, 4), the optimized Fermi-level configurations effectively suppress unwanted diffraction orders, concentrating the transmitted power into the targeted 1$^{st}$ order diffraction angle. The residual efficiency in higher-order channels remains negligible, confirming the effectiveness of the gradient-descent optimization in directing the transmitted light toward the intended steering angle.

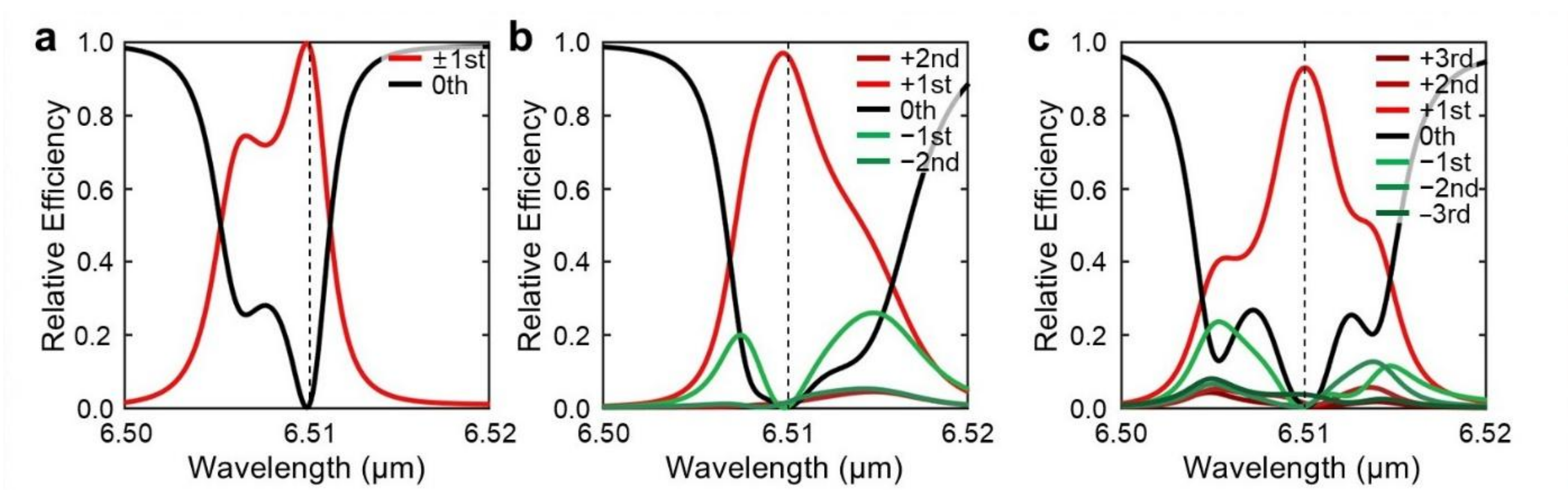

**Figure S15.** Relative efficiency of active beam-steering metasurface shown in Fig. 4. Relative efficiency of each diffraction order at the resonant wavelength when (a) two, (b) three, and (c) four Fermi levels are applied.

## S8. Verification of single-resonance system

We verify six-port single-resonance system using S4 RCWA software. At $E_F = 0.6$ eV, the pole lies above the real axis, and a branch cut connecting the zero and pole crosses the real frequency axis, yielding a $2\pi$ phase coverage in the frequency domain[5] (Figure S16a). At $E_F$ = CNP, the loss increases as the zero moves toward the real axis, confirming transmission null (Figure S16a).

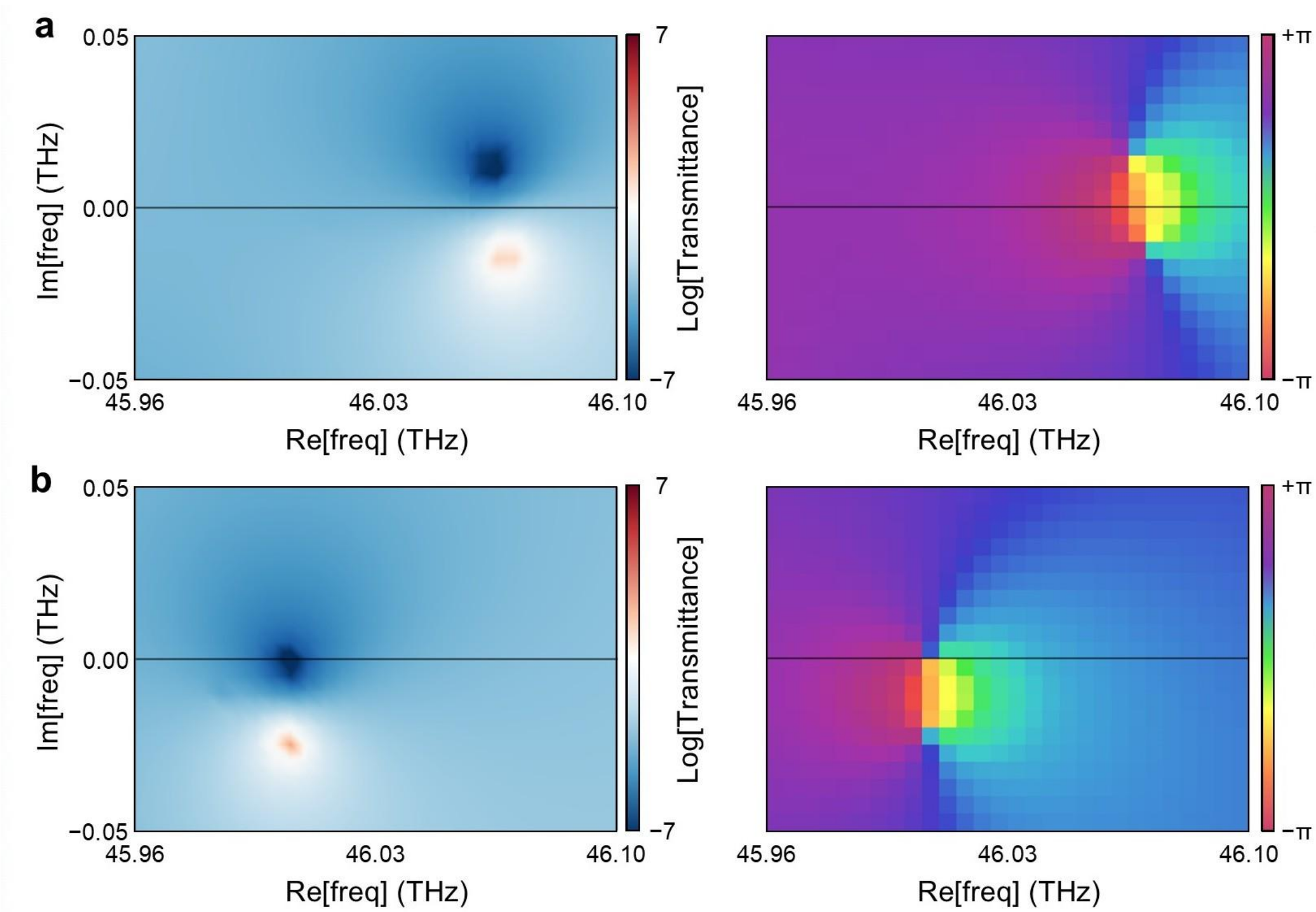

**Figure S16.** Single-resonance verification via complex frequency analysis. (a, b) Complex frequency plots of $\log(|t|^2)$ (left) and $\arg(t)$ (right) for the optimized structure shown in Figure 4 of the main text ($T = 310$ K) at (a) $E_F = 0.6$ eV and (b) $E_F$ = CNP. The presence of a single zero–pole pair near the real frequency axis confirms that the metasurface operates in the single-resonance regime across the relevant Fermi-level range. At $E_F = 0.6$ eV, the pole lies above the real frequency axis, corresponding to a $2\pi$ phase coverage in the frequency domain. At $E_F$ = CNP, the increased graphene-induced loss shifts the pole below the real axis, producing a transmission null that enables unity amplitude modulation

## S9. RCWA convergence verification with FEM verification

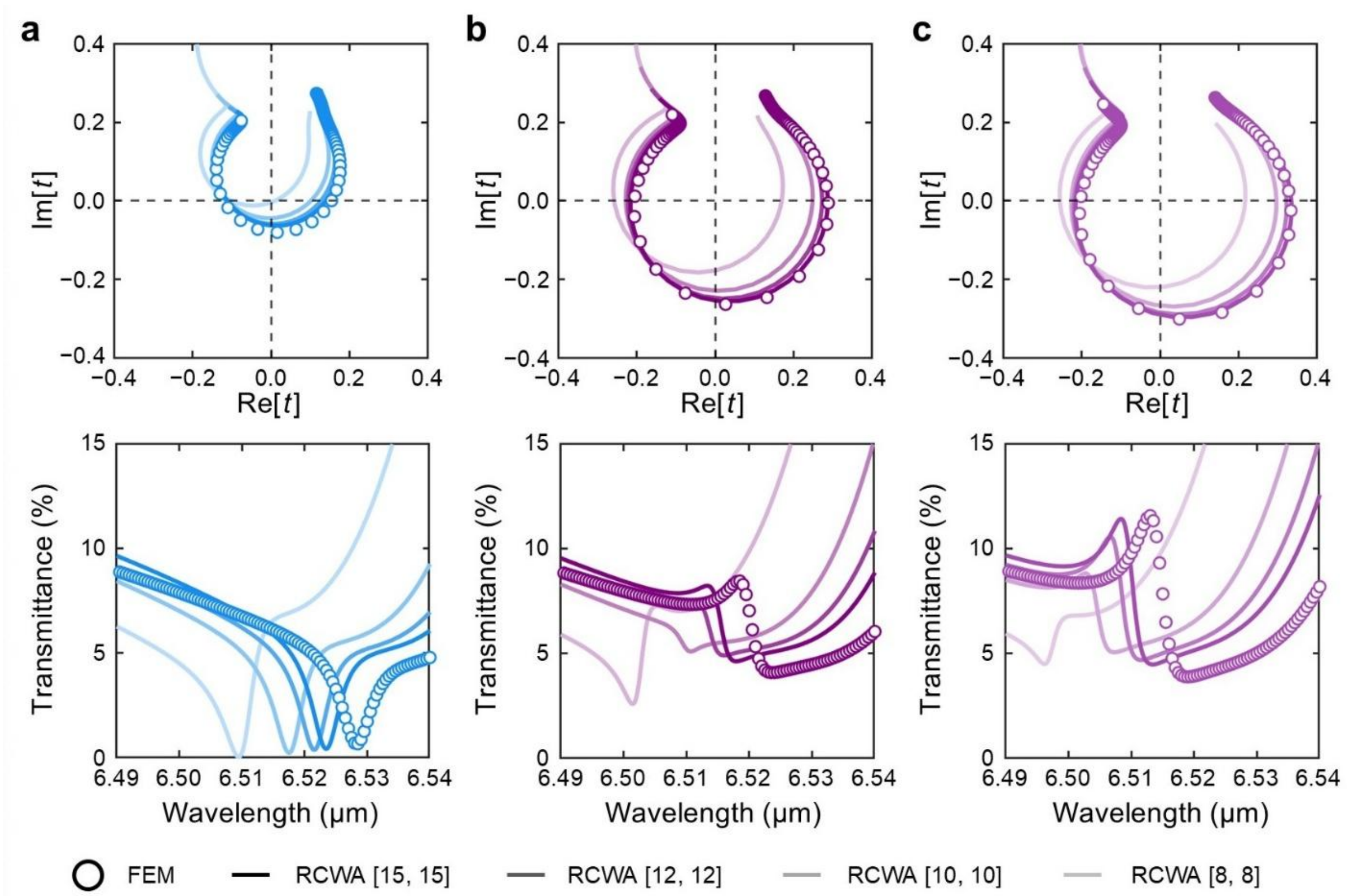

**Figure S17.** Simulation verification of the single unit cell. Complex transmission coefficient trajectories (top) and transmittance spectra (bottom) of the optimized structure in Figure 4 of the main text, computed at $T = 310$ K for (a) $E_F = 0.1$ eV, (b) 0.6 eV, and (c) 1.0 eV, computed using RCWA with different Fourier orders ([8, 8], [10, 10], [12, 12], [15, 15]) and validated against FEM simulation (COMSOL Multiphysics), showing convergence with increasing Fourier order and excellent agreement with the FEM results

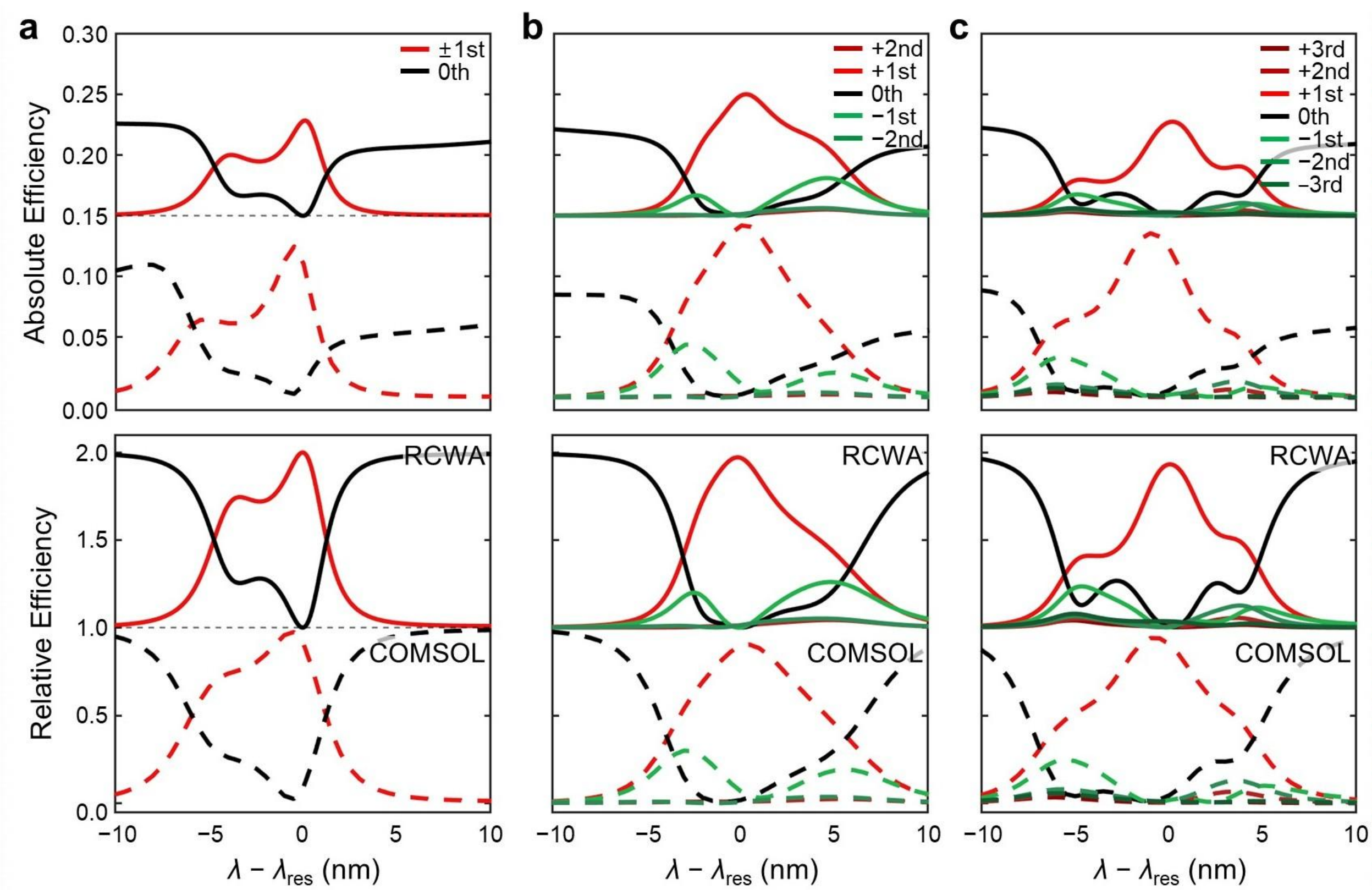

**Figure S18.** Simulation verification of the phase gradient metasurface. (a–c) Simulated absolute (top row) and relative (bottom row) diffraction efficiencies of the gradient metasurface configured with (a) two-level ($N$ = 2), (b) three-level ($N$ = 3), and (c) four-level ($N$ = 4) phase quantization. The efficiency spectra are plotted as a function of wavelength detuning ($\lambda$ - $\lambda_{res}$). Solid lines represent the optical responses calculated using RCWA, while the dashed lines denote the cross-validation results obtained via the FEM simulation (COMSOL Multiphysics), demonstrating excellent numerical agreement between the two independent solvers.